\documentclass[aps,prd,10pt,notitlepage,nofootinbib,superscriptaddress,showkeys,showpacs]{revtex4-1}
\pdfoutput=1
\usepackage{amstext,amsmath,amssymb,amsfonts,bbm,euscript,color,dsfont}
\usepackage[latin1]{inputenc}
\usepackage{epsfig}
\usepackage{hyperref}
\usepackage{amsthm}
\usepackage{subfigure}
\usepackage{color}
\usepackage{multirow}
\usepackage{tikz}

\usetikzlibrary{arrows,backgrounds}

\usepackage{verbatim}

\usepackage{graphicx}

\usepackage{latexsym}

\newcommand{\bea}{\begin{eqnarray}}	
\newcommand{\eea}{\end{eqnarray}}
\newcommand{\be}{\begin{equation}}	
\newcommand{\ee}{\end{equation}}
\newcommand{\beq}{\begin{equation}}	
\newcommand{\eeq}{\end{equation}}

\newcommand{\Z}{{\mathbb Z}}
\newcommand{\C}{{\mathbb C}}

\newcommand{\dd}{{\textrm{d}}}

\def\R{\relax\ifmmode {\mathbb R}  \else${\mathbb R}$\fi}
\def\C{\relax\ifmmode {\mathbb C}  \else${\mathbb C}$\fi}
\def\Z{\relax\ifmmode {\mathbb Z}  \else${\mathbb Z}$\fi}
\def\N{\relax\ifmmode {\mathbb N}  \else${\mathbb N}$\fi}
\def\I{\relax\ifmmode {\mathbb I}  \else${\mathbb I}$\fi}

\begin{document}

\title{Renormalizability  of the Refined Gribov-Zwanziger action in the linear covariant gauges}  

%\begin{comment}

\author{M.~A.~L.~Capri}\email{caprimarcio@gmail.com}

\affiliation{UERJ $-$ Universidade do Estado do Rio de Janeiro,\\
Departamento de F\'isica Te\'orica, Rua S\~ao Francisco Xavier 524,\\
20550-013, Maracan\~a, Rio de Janeiro, Brasil}

\author{D.~Fiorentini}\email{diegodiorentinia@gmail.com}

\affiliation{UERJ $-$ Universidade do Estado do Rio de Janeiro,\\
Departamento de F\'isica Te\'orica, Rua S\~ao Francisco Xavier 524,\\
20550-013, Maracan\~a, Rio de Janeiro, Brasil}

\author{A.~D.~Pereira}\email{aduarte@if.uff.br}

\affiliation{UFF $-$ Universidade Federal Fluminense,\\
Instituto de F\'{\i}sica, Campus da Praia Vermelha,\\
Avenida General Milton Tavares de Souza s/n,\\ 
24210-346, Niter\'oi, RJ, Brasil}

\author{S.~P.~Sorella}\email{silvio.sorella@gmail.com}

\affiliation{UERJ $-$ Universidade do Estado do Rio de Janeiro,\\
Departamento de F\'isica Te\'orica, Rua S\~ao Francisco Xavier 524,\\
20550-013, Maracan\~a, Rio de Janeiro, Brasil}

%\end{comment}

\begin{abstract}
The Refined Gribov-Zwanziger framework takes into account the existence of equivalent gauge field configurations in the  gauge-fixing  quantization procedure of Euclidean Yang-Mills theories. Recently, this setup was extended to the family of linear covariant gauges giving rise to a local and BRST-invariant action. In this paper, we give an algebraic proof of the renormalizability of the resulting action to all orders in perturbation theory.

%\
 
%\noindent Pacs numbers: ...\\  
%\noindent Key words: ... \\ 
%\noindent Report numbers: xxxxx

\end{abstract}

\maketitle

\tableofcontents

%%%%%%%%%%%%%%%%%%%%%%%%%%%%%%%%%%%%%%%%%%%%%%%%%%%%%%%%%%%%%%
\section{Introduction}
%%%%%%%%%%%%%%%%%%%%%%%%%%%%%%%%%%%%%%%%%%%%%%%%%%%%%%%%%%%%%%

A general feature of the continuum formulation of non-Abelian gauge theories is the gauge-fixing procedure. As is widely known, this can be achieved through the so-called Faddeev-Popov method which gives consistent results within the perturbative treatment of gauge theories. Nevertheless, as was shown by Gribov in \cite{Gribov:1977wm}, the Faddeev-Popov procedure relies on some hypothesis which are not well grounded as long as one goes away from the perturbative regime. The problem arises from the fact that a gauge-fixing condition is not enough to fix completely the gauge freedom, allowing for gauge equivalent configurations, the so-called Gribov copies, even after imposing the gauge-fixing condition. It was soon realized that this is not a particular problem of some specific gauge-fixing, but an intrinsic problem related to the non-trivial geometrical structure non-Abelian gauge theories, see \cite{Singer:1978dk}. For a pedagogical introduction to the Gribov problem, we refer to \cite{Sobreiro:2005ec,Vandersickel:2012tz,Vandersickel:2011zc,Pereira:2016inn}. \\\\In the Landau gauge, the gauge-fixing condition is expressed as

\begin{equation}
\partial_\mu A^a_\mu = 0\,.
\label{intro1}
\end{equation}  
Such a condition would be ideal if any gauge equivalent configuration $A'^a_\mu$, connected through $A^a_\mu$ via a gauge transformation, would  not satisfy \eqref{intro1}. For concreteness, one can assume that $A^a_\mu$ and $A'^a_\mu$ are connected via an infinitesimal gauge transformation. Hence,

\begin{equation}
A'^a_\mu = A^a_\mu - D^{ab}_{\mu}\theta^b\,,\,\,\,\,\,\partial_\mu A'^{a}_{\mu} = 0\,\,\,\,\Rightarrow\,\,\,\, -\partial_\mu D^{ab}_{\mu}\theta^b=0\,,
\label{intro2}
\end{equation}
with $\theta^a$ an infinitesimal gauge parameter and $D^{ab}_{\mu}\equiv \delta^{ab}\partial_\mu - gf^{abc}A^{c}_{\mu}$, the covariant derivative in the adjoint representation of the gauge group\footnote{We consider $SU(N)$ gauge theories in four Euclidean dimensions.}. Eq.\eqref{intro2} reveals that the configuration $A'^a_\mu$ satisfies the gauge condition \eqref{intro1}, {\it i.e.} it is a Gribov copy of $A^a_\mu$,  if the operator $-\partial_\mu D^{ab}_\mu$ develops zero-modes. In \cite{Gribov:1977wm}, it was proven that such zero-modes exist and explicit examples were constructed. As a consequence, even after the gauge-fixing procedure, a residual gauge redundancy remains or, in other words, the gauge-fixing is not ideal. Therefore, the Faddeev-Popov procedure should not be strictly applied as it stands, requiring  an improvement at the non-perturbative level. \\\\Already in \cite{Gribov:1977wm}, Gribov proposed that, besides the standard gauge-fixing, a further constraint should be imposed to the path integral: the functional measure should be restricted to a region free from zero-modes of the Faddeev-Popov operator $-\partial_\mu D^{ab}_\mu$. Such a region, known as the \textit{Gribov region} $\Omega_L$, is defined as

\begin{equation}
\Omega_L = \left\{\,A^a_\mu\,, \,\,\partial_\mu A^a_\mu = 0\,|\,-\partial_\mu D^{ab}_\mu >0\right\}\,,
\label{intro3}
\end{equation}
where the subscript $L$ means that one is referring to the Landau gauge. It is important to point out here that, in  this gauge, the Faddeev-Popov operator  $-\partial_\mu D^{ab}_\mu$, is hermitian, implying that its eigenvalues are real. This property  allows for a meaningful requirement of the positivity of the Faddeev-Popov operator. Moreover, it has been proven in \cite{Dell'Antonio:1991xt} that the region $\Omega_L$ enjoys a set of remarkable features : \textit{i)} It is bounded in all directions in field space; \textit{ii)} The trivial perturbative vacuum $A^a_\mu = 0$ belongs to $\Omega_L$; \textit{iii)} It is convex; \textit{iv)} All gauge orbits cross $\Omega_L$ at least once. From \eqref{intro2} and \eqref{intro3}, it is clear that the Gribov region is free from infinitesimal Gribov copies.  Unfortunately, it still contains copies generated by large gauge transformations and one should look for a region truly free from Gribov copies, known as the \textit{fundamental modular region}, see \cite{vanBaal:1991zw}. Ideally, one should restrict the path integral to the fundamental modular region rather than the Gribov region. Nevertheless, till now, a practical operational  way to restrict the functional integral to  the fundamental modular region has not yet been achieved. We stick therefore to the Gribov region  $\Omega_L$. \\\\Formally, Gribov's proposal can  written as

\begin{equation}
\EuScript{Z} = \int_{\Omega_L} \left[\EuScript{D}\Phi\right]\mathrm{e}^{-(S_{\mathrm{YM}}+S_{\mathrm{FP}})}\,,
\label{intro4}
\end{equation}
with

\begin{equation}
S_{\mathrm{YM}} = \frac{1}{4}\int \dd^4x~ F^{a}_{\mu\nu}F^{a}_{\mu\nu}\,,\,\,\,\,\qquad \,\,\,\,\,S_{\mathrm{FP}}=\int \dd^4x\left(ib^a\partial_{\mu}A^a_\mu + \bar{c}^a\partial_\mu D^{ab}_\mu c^b\right)\,,
\label{intro5}
\end{equation}
where $b^a$, $\bar{c}^a$, $c^a$ are the auxiliary Nakanishi-Lautrup field, the Faddeev-Popov antighost and ghost fields, respectively, and $\Phi$ is a shorthand notation for all the fields of the theory. The field strength $F^a_{\mu\nu}$ is given by 

\begin{equation}
F^{a}_{\mu\nu} = \partial_{\mu}A^a_\nu - \partial_{\nu}A^a_\mu + gf^{abc}A^{b}_{\mu}A^{c}_{\nu}\,.
\label{intro6}
\end{equation} 
The implementation of the restriction to the region $\Omega_L$, as expressed in eq.\eqref{intro4},  was worked out at leading order by Gribov in \cite{Gribov:1977wm} and to all orders by Zwanziger in \cite{Zwanziger:1989mf}. Although the procedures pursued are different, their equivalence was established to all orders in \cite{Capri:2012wx}. Effectively, the restriction to $\Omega_L$ is achieved through the following modification of the original Faddeev-Popov path integral,

\begin{equation}
\EuScript{Z} = \int_{\Omega_L} \left[\EuScript{D}\Phi\right]\mathrm{e}^{-(S_{\mathrm{YM}}+S_{\mathrm{FP}})}  = \int \left[\EuScript{D}\Phi\right]\mathrm{e}^{-(S_{\mathrm{YM}}
+S_{\mathrm{FP}}+\gamma^4H(A)-4V\gamma^4(N^2-1))}\,,
\label{intro7}
\end{equation}
where

\begin{equation}
H(A)=g^2\int \dd^4x\dd^4y~f^{abc}A^{b}_{\mu}(x)\left[\EuScript{M}^{-1}\right]^{ad}(x,y)f^{dec}A^{e}_{\mu}(y)\,,
\label{intro8}
\end{equation}
is the so-called \textit{horizon function}, with $\EuScript{M}^{ab}=-\partial_\mu D^{ab}_\mu$, $V$ is the spacetime volume and $N$ is the number of colors. The parameter $\gamma$ has mass dimension one and is known as the Gribov parameter. It is not free, but determined through a gap equation,

\begin{equation}
\langle H(A)\rangle=4V(N^2-1)\,,
\label{intro9}
\end{equation}
with $\langle \ldots \rangle$ taken with the functional measure defined in \eqref{intro7}. As is evident from \eqref{intro8}, the horizon function is non-local, giving rise to  a non-local action which, thanks to the aforementioned properties of the Gribov region $\Omega_L$\footnote{We underline here that the important property that all gauge orbits cross $\Omega_L$ at least once, gives a well defined support to original Gribov's proposal of restricting the domain of integration in the path integral to the region $\Omega_L$, although $\Omega_L$ itself is not free from Gribov copies.}, takes into account the existence of a huge set of Gribov copies. Remarkably, such an action can be localized through the introduction of auxiliary fields viz. a pair of bosonic ones $(\bar{\varphi},\varphi)^{ab}_\mu$ and a pair of anticommuting fields $(\bar{\omega},\omega)^{ab}_\mu$. In the Landau gauge, the local action takes the following form 

\begin{equation}
S_{\mathrm{GZ}}=S_{\mathrm{YM}}+S_{\mathrm{FP}}-\int \dd^4x\left(\bar{\varphi}^{ac}_{\mu}\EuScript{M}^{ab}{\varphi}^{bc}_{\mu}-\bar{\omega}^{ac}_\mu\EuScript{M}^{ab}\omega^{bc}_\mu+gf^{adl}\bar{\omega}^{ac}_\mu\partial_\nu\left(\varphi^{lc}_\mu D^{de}_\nu c^e\right)\right)-\gamma^{2}\int \dd^4x~gf^{abc}A^{a}_{\mu}(\varphi+\bar{\varphi})^{bc}_\mu\,,
\label{intro10}
\end{equation}
and is known as the Gribov-Zwanziger action. It is easy to check that,  upon integration of the auxiliary fields, the non-local action \eqref{intro7} is recovered. This action is renormalizable to all orders in perturbation theory \cite{Zwanziger:1989mf} and provides a local framework which, thanks to the aforementioned properties of the Gribov region $\Omega_L$,  takes into account the existence of a huge set of Gribov copies in the Landau gauge. \\\\Nevertheless, It was realized  in \cite{Dudal:2008sp} that the restriction to the Gribov region leads to additional non-perturbative  instabilities giving rise to the formation of dimension-two condensates. In particular, it was shown that the condensates $\langle A^{a}_{\mu} A^{a}_{\mu}\rangle$ and $\langle\bar{\varphi}^{ab}_{\mu}\varphi^{ab}_{\mu}-\bar{\omega}^{ab}_{\mu} \omega^{ab}_{\mu}\rangle$ are non-vanishing already at one-loop order, besides being free of ultraviolet divergences, see also \cite{Dudal:2007cw,Dudal:2011gd,Gracey:2010cg}. \\\\Taking into account the existence of such condensates from the beginning, gives rise to the so-called Refined Gribov-Zwanziger (RGZ) action, which is expressed as

\begin{equation}
S_{\mathrm{RGZ}}=S_{\mathrm{GZ}}+\frac{m^2}{2}\int\dd^4x~A^{a}_{\mu}A^{a}_{\mu}-M^2\int \dd^4x\left(\bar{\varphi}^{ab}_{\mu}\varphi^{ab}_{\mu}-\bar{\omega}^{ab}_{\mu}\omega^{ab}_{\mu}\right)\,,
\label{intro11}
\end{equation}  
where the mass parameters $m$ and $M$ are dynamically determined by their own gap equations, see \cite{Dudal:2011gd}. The RGZ action enjoys many interesting properties. In particular, the tree-level gluon propagator reads

\begin{equation} 
\langle A^a_\mu(k) A^b_\nu(-k) \rangle = \delta^{ab} \frac{k^2 + M^2}{(k^2+m^2)(k^2+M^2) + 2g^2N \gamma^4} \left( \delta_{\mu\nu} - \frac{k_\mu k_\nu}{k^2} \right)\,, 
\label{intro12} 
\end{equation}
from which one sees that it attains a finite value at $k=0$. Such a massive/decoupling behavior is in agreement with the most recent lattice data as well as with functional and effective methods, \cite{Maas:2011se,Cucchieri:2007rg,Aguilar:2008xm,Fischer:2008uz,Cucchieri:2009zt,Tissier:2010ts,Cyrol:2016tym,Reinosa:2017qtf}. As observed in \cite{Dudal:2008xd}, in $d=2$ the refinement does not occur due to the existence of infrared divergences which prevent the formation of the condensates. As a consequence, in $d=2$ the gluon propagator vanishes at zero momentum, giving rise to the so-called scaling solution \cite{Alkofer:2000wg}. In $d=3$ \cite{Dudal:2008rm} as well as $d>4$, \cite{Guimaraes:2016okb}, the massive/decoupling behavior persists. Such different behavior for $d=2$ and $d>3$ is also observed in lattice simulations, see \cite{Cucchieri:2011ig}. Let us  emphasize that the propagator \eqref{intro12} displays positivity violation \cite{Dudal:2008sp,Cucchieri:2011ig}, a fact that does not allow to interpret gluons as excitations of the physical spectrum of the theory, giving thus a strong signal of color confinement.  \\\\The Gribov-Zwanziger formulation has been object of intensive investigation in recent years, with particular emphasis on the establishment of its BRST invariance, see, for instance,   \cite{Maggiore:1993wq,Serreau:2012cg,Serreau:2015yna,Capri:2014bsa,Dudal:2009xh,Sorella:2009vt,Baulieu:2008fy,Capri:2010hb,Dudal:2012sb,Dudal:2014rxa,Pereira:2013aza,Pereira:2014apa,Lavrov:2011wb,Lavrov:2013boa,Moshin:2015gsa,Schaden:2014bea,Cucchieri:2014via} for previous attempts. Recently, in  \cite{Capri:2015ixa}, the existence of a manifest  exact BRST invariance of the  Gribov-Zwanziger action and of its refined version has been achieved through  the use of the non-local, transverse and gauge-invariant field\footnote{We refer to Appendix.~A of \cite{Capri:2015ixa} for the details of the construction of $A^{h,a}_\mu$.} $A^{h,a}_{\mu}$, introduced in \cite{Zwanziger:1990tn,Lavelle:1995ty}. Explicitly, $A^{h,a}_\mu$ can be written as an infinite non-local series,   given by\footnote{We employ a matrix notation, as described in Appendix.~A of \cite{Capri:2015ixa}.}

\begin{equation}
A^{h}_{\mu}=\left(\delta_{\mu\nu}-\frac{\partial_{\mu}\partial_{\nu}}{\partial^2}\right)\left(A_{\nu}-ig\left[\frac{1}{\partial^2}\partial A,A_\nu\right]+\frac{ig}{2}\left[\frac{1}{\partial^2}\partial A,\partial_{\nu}\frac{1}{\partial^2}\partial A\right]+\mathcal{O}(A^3)\right)\,,
\label{intro13}
\end{equation}
with $A^h_\mu$ being BRST-invariant. 

\begin{equation}
sA^h_\mu = 0\,, \qquad s A^a_\mu = - D^{ab}_{\mu}(A) c^b \;.
\label{intro14}
\end{equation}
When written in terms of $A^h_\mu$, the RGZ action in the Landau gauge can be expressed as \cite{Capri:2015ixa}:  

\begin{eqnarray}
\tilde{S}^{L}_{\mathrm{RGZ}}&=&S_{\mathrm{YM}}+S_{\mathrm{FP}}-\int \dd^4x\left(\bar{\varphi}^{ac}_{\mu}\EuScript{M}^{ab}(A^h){\varphi}^{bc}_{\mu}-\bar{\omega}^{ac}_\mu\EuScript{M}^{ab}(A^h)\omega^{bc}_\mu\right)-\gamma^{2}\int \dd^4x~gf^{abc}A^{h,a}_{\mu}(\varphi+\bar{\varphi})^{bc}_\mu\nonumber\\
&+&\frac{m^2}{2}\int\dd^4x~A^{h,a}_{\mu}A^{h,a}_{\mu}-M^2\int \dd^4x\left(\bar{\varphi}^{ab}_{\mu}\varphi^{ab}_{\mu}-\bar{\omega}^{ab}_{\mu}\omega^{ab}_{\mu}\right)\,,
\label{intro15}
\end{eqnarray}
with

\begin{equation}
\EuScript{M}^{ab}(A^h) = -\delta^{ab}\partial^2+gf^{abc}A^{h,c}_{\mu}\partial_\mu\,,\,\,\,\,\mathrm{with}\,\,\,\, \partial_{\mu}A^{h,a}_{\mu} = 0\,.
\label{intro16}
\end{equation}
Despite the presence of the Zwanziger's localizing fields, the action \eqref{intro15} is still non-local due to the presence of the non-local field $A^{h,a}_{\mu}$, see eq.\eqref{intro13}. Notwithstanding, a localization of such an action was introduced in \cite{Capri:2016aqq} by means of the use of a Stueckelberg-like field $\xi^a$. More specifically, following  \cite{Capri:2016aqq}, the non-local expression \eqref{intro13}  is re-expressed in a local way as 

\begin{equation}
A^{h}_{\mu} = h^\dagger A_\mu h +\frac{i}{g}h^\dagger \partial_\mu h\,,
\label{intro17}
\end{equation}
where

\begin{equation}
h = \mathrm{e}^{ig\xi^a T^a}\,,
\label{intro18}
\end{equation}
with $\xi$ being an auxiliary Stueckelberg field. In addition, one imposes the transversality condition 
\begin{equation}
\partial_\mu A^{h}_\mu = 0 \;. \label{trah}
\end{equation} 
When solved iteratively for the Stueckelberg field, equations \eqref{intro17},\eqref{trah} give back the non-local expression \eqref{intro13}, see \cite{Capri:2016aqq} for the details. \\\\Therefore, for the local and BRST-invariant RGZ action in the Landau gauge one gets \cite{Capri:2016aqq}:

\begin{eqnarray}
S^{L}_{\mathrm{RGZ}}&=&S_{\mathrm{YM}}+S_{\mathrm{FP}}-\int \dd^4x\left(\bar{\varphi}^{ac}_{\mu}\EuScript{M}^{ab}(A^h){\varphi}^{bc}_{\mu}-\bar{\omega}^{ac}_\mu\EuScript{M}^{ab}(A^h)\omega^{bc}_\mu\right)-\gamma^{2}\int \dd^4x~gf^{abc}(A^h)^a_{\mu}(\varphi+\bar{\varphi})^{bc}_\mu\nonumber\\
&+&\frac{m^2}{2}\int\dd^4x~(A^h)^a_{\mu}(A^h)^a_{\mu}-M^2\int \dd^4x\left(\bar{\varphi}^{ab}_{\mu}\varphi^{ab}_{\mu}-\bar{\omega}^{ab}_{\mu}\omega^{ab}_{\mu}\right)+\int \dd^4x~\tau^a \partial_\mu (A^h)^a_\mu - \int \dd^4x~\bar{\eta}^a \EuScript{M}^{ab}(A^h)\eta^b\,,
\label{intro19}
\end{eqnarray}  
with $\tau^a$ being a Lagrange multiplier needed to  imposes the transversality of $A^h_\mu$, eq.\eqref{trah}. The fields  $(\bar{\eta}^a,\eta^a)$ are a pair of ghosts needed to take into account the Jacobian\footnote{In \cite{Capri:2016aqq} the ghosts $(\bar{\eta},\eta)^a$ were not introduced. Although this term does not alter the results of \cite{Capri:2016aqq}, such a term is needed in order to prove the equivalence between the  local and non-local and formulations. We acknowledge U. Reinosa, J. Serreau, M. Tissier and N. Wschebor for discussions on this issue.} arising from the transversality constraint \eqref{trah}, $\partial_\mu A^{h}_\mu = 0$. We remind the reader to Appendix.~A for the proof of the equivalence between the RGZ actions given in eqs.\eqref{intro19} and  \eqref{intro11}. As a consequence, the functional integral for the local and BRST-invariant RGZ action in the Landau gauge is written as

\begin{equation}
\EuScript{Z} = \int \left[\EuScript{D}\mu\right]\mathrm{e}^{-S_{\mathrm{RGZ}}+4V\gamma^4(N^2-1)}\,,
\label{intro20}
\end{equation} 
where

\begin{equation}
\left[\EuScript{D}\mu\right] = \left[\EuScript{D}A\right]\left[\EuScript{D}b\right]\left[\EuScript{D}\bar{c}\right]\left[\EuScript{D}c\right]\left[\EuScript{D}\bar{\varphi}\right]\left[\EuScript{D}\varphi\right]\left[\EuScript{D}\bar{\omega}\right]\left[\EuScript{D}\omega\right]\left[\EuScript{D}\tau\right]\left[\EuScript{D}\xi\right]\left[\EuScript{D}\bar{\eta}\right]\left[\EuScript{D}\eta\right]\,.
\label{intro21}
\end{equation}
Explicitly, the nilpotent BRST transformations which leave the action \eqref{intro19} invariant are

\begin{align}
sA^{a}_{\mu}&=-D^{ab}_{\mu}c^b\,,     &&sc^a=\frac{g}{2}f^{abc}c^bc^c\,, \nonumber\\
s\bar{c}^a&= ib^{a}\,,     &&sb^{a}=0\,, \nonumber\\
s\varphi^{ab}_{\mu}&=0\,,   &&s\omega^{ab}_{\mu}=0\,, \nonumber\\
s\bar{\omega}^{ab}_{\mu}&=0\,,         &&s\bar{\varphi}^{ab}_{\mu}=0\,,\nonumber \\
s h^{ij}& = -ig c^a (T^a)^{ik} h^{kj}  \;, && sA^{h,a}_\mu =0\,,   \nonumber \\
s\tau^a& =0\,, && s\bar{\eta}^a=0 \,, \nonumber \\
s\eta^a &= 0\,, && s^2 = 0\,.  \label{intro22}
\end{align}
The BRST transformation of the Stueckelberg field $\xi^a$ can be obtained iteratively from the transformation of $h^{ij}$, {\it i.e.} $(sh^{ij}=-ig c^a (T^a)^{ik} h^{kj})$, yielding 
\begin{equation}
s\xi^a = g^{ab}(\xi)c^b\,, 
\end{equation} 
where $g^{ab}(\xi)$ is a power series in $\xi^a$, namely 
\begin{equation}
g^{ab}(\xi) =  - \delta^{ab} + \frac{g}{2} f^{abc} \xi^c - \frac{g^2}{12} f^{amr} f^{mbq}  \xi^q \xi^r + O(g^3)\,.
\label{intro22a}
\end{equation}
Having a local and BRST invariant setup which takes into account the existence of Gribov copies in the Landau gauge, it is natural to look for an extension of such a framework to different gauges. A natural generalization are the so-called linear covariant gauges, where the gauge condition reads

\begin{equation}
\partial_\mu A^a_\mu = -i \alpha b^a\,,
\label{intro23}
\end{equation}
with $\alpha$ a non-negative gauge parameter. Although this class of gauges preserves the linearity of the Landau gauge, it introduces a gauge parameter $\alpha$ as well as a longitudinal sector for the gluon fields. In a series of papers \cite{Capri:2015ixa,Capri:2016aqq,Capri:2015nzw,Capri:2016gut,Sobreiro:2005vn,Capri:2015pja,Capri:2016aif,Capri:2017abz}, the development of the RGZ action to linear covariant gauges was worked out. Besides being interesting by its own as a further development of the RGZ framework, recent studies within analytic as well as numerical lattice approaches to non-perturbative Yang-Mills theories started employing the linear covariant gauges \cite{Cucchieri:2009kk,Cucchieri:2011pp,Aguilar:2015nqa,Huber:2015ria,Bicudo:2015rma,Aguilar:2016ock}.  Hence, an interplay between these different  approaches, as performed in the Landau gauge,  becomes possible, leading to a deeper understanding of the behavior of the correlation functions in the non-perturbative infrared region in this class of gauges. \\\\In this paper we pursue the study of the RGZ formulation initiated in \cite{Capri:2015ixa,Capri:2016aqq,Capri:2015nzw,Capri:2016gut,Sobreiro:2005vn,Capri:2015pja,Capri:2016aif,Capri:2017abz}, by addressing  the issue of  the renormalizability properties of the local and BRST invariant RGZ action in the linear covariant gauges. In particular, we prove, using the algebraic renormalization setup \cite{Piguet:1995er},  the all orders renormalizability of the RGZ framework in the linear covariant gauge, a topic which was still lacking in our previous studies. \\\\The structure of the paper is the following: In Sect.~\ref{RGZLCGreview} we give a brief overview of the construction of the RGZ action in the linear covariant gauges. Subsequently, in Sect.~\ref{Renpreli} we identify the classical complete action which will be the starting point for the algebraic renormalization analysis. In Sect.~\ref{Wardid} we give the formal proof of the all order renormalizability of the RGZ action in the linear covariant gauges.  Finally, we collect our conclusions. Additional material clarifying some specific technical points of this work are collected in the appendix.

%%%%%%%%%%%%%%%%%%%%%%%%%%%%%%%%%%%%%%%%%%%%%%%%%%%%%%%%%%%%%%%%%%%%%%%%%%%%
\section{The Refined Gribov-Zwanziger action in the linear covariant gauges} \label{RGZLCGreview}
%%%%%%%%%%%%%%%%%%%%%%%%%%%%%%%%%%%%%%%%%%%%%%%%%%%%%%%%%%%%%%%%%%%%%%%%%%%%

Within the standard Faddeev-Popov framework, the gauge-fixed Yang-Mills action in the linear covariant gauges reads

\begin{equation}
S^{\mathrm{FP}}_{\mathrm{LCG}} = S_{\mathrm{YM}} + \int \dd^4x\left(ib^a \partial_{\mu}A^{a}_{\mu}+ \frac{\alpha}{2}b^a b^a + \bar{c}^a\partial_{\mu}D^{ab}_{\mu}c^b\right) = S_{\mathrm{YM}} + \int d^4x\; s\left( {\bar c^a} \partial_\mu A^a_\mu - \frac{i\alpha}{2} {\bar c^a} b^a \right) \,,
\label{rgzlcg1}
\end{equation}
where the gauge parameter $\alpha$ is non-negative. The particular case $\alpha = 0$ is the Landau gauge. Very much close to the case of the Landau gauge, infinitesimal Gribov copies arise as long as the Faddeev-Popov operator $-\partial_\mu D^{ab}_\mu$ develops zero-modes. Nevertheless, for non-vanishing $\alpha$, such an operator is not Hermitean, a feature that jeopardizes the standard Gribov-Zwanziger analysis for the removal of such zero-modes from the path integral measure. This problem has been a great challenge in dealing with Gribov copies for generic values of $\alpha$. A first attempt to face this issue was to consider the operator $-\partial_\mu D^{ab}_\mu$ projected onto the transverse component of the gauge field, see \cite{Sobreiro:2005vn,Capri:2015pja,Capri:2016aif}. In this case, the resulting projected operator is Hermitean and the standard procedure for the construction of a horizon-like function is available. This setup was worked out in \cite{Sobreiro:2005vn,Capri:2015pja} and its  renormalizability was analyzed in details in \cite{Capri:2016aif}. Nonetheless, it exhibits drawbacks: \textit{i)} the limit $\alpha = 0$ does not fully recover the standard (R)GZ action in the Landau gauge and \textit{ii)} it breaks BRST symmetry softly, a feature that obscures  the control of the gauge parameter  independence of the correlation functions of  gauge invariant quantities. These difficulties have been overcome by the construction of a BRST invariant formulation of the RGZ action in the Landau gauge which,  as discussed in \cite{Capri:2015ixa}, naturally leads to a BRST invariant formulation of the linear covariant gauges.  \\\\Following \cite{Capri:2015ixa}, the local and BRST-invariant RGZ action in the linear covariant gauges is written as

\begin{eqnarray}
S^{\mathrm{LCG}}_{\mathrm{RGZ}}&=& S_{\mathrm{LCG}}^{\mathrm{FP}}-\int \dd^4x\left(\bar{\varphi}^{ac}_{\mu}\EuScript{M}^{ab}(A^h){\varphi}^{bc}_{\mu}-\bar{\omega}^{ac}_\mu\EuScript{M}^{ab}(A^h)\omega^{bc}_\mu\right)-\gamma^{2}\int \dd^4x~gf^{abc}(A^h)^a_{\mu}(\varphi+\bar{\varphi})^{bc}_\mu\nonumber\\
&+&\frac{m^2}{2}\int\dd^4x~(A^h)^a_{\mu}(A^h)^a_{\mu}-M^2\int \dd^4x\left(\bar{\varphi}^{ab}_{\mu}\varphi^{ab}_{\mu}-\bar{\omega}^{ab}_{\mu}\omega^{ab}_{\mu}\right)+\int \dd^4x~\tau^a \partial_\mu (A^h)^a_\mu - \int \dd^4x~\bar{\eta}^a \EuScript{M}^{ab}(A^h)\eta^b\,.
\label{rgzlcg2}
\end{eqnarray}
This action is invariant under the nilpotent BRST transformations (\ref{intro22}). As such, the following properties can be shown to hold, see \cite{Capri:2016aqq,Capri:2015nzw,Capri:2016gut}: 

\begin{itemize}
\item Correlation functions of gauge-invariant quantities are $\alpha$-independent;

\item The mass parameters $(\gamma,m,M)$ are independent from the gauge parameter $\alpha$ and, as a consequence, can enter physical quantities;

\item The longitudinal part of the gluon propagator is exact and equal to the tree-level result;

\item The pole mass of the transverse component of the gluon propagator is independent from $\alpha$.

\item The action \eqref{rgzlcg2} effectively implements the restriction of the path integral to the functional region $\Sigma$ defined as

\begin{equation}
\Sigma = \left\{\,A^a_\mu\,, \,\,\partial_\mu A^a_\mu = \alpha i b^a\,\Big|\,-\partial_\mu D^{ab}_\mu (A^h) >0\right\}\,.
\label{rgzlcg2a}
\end{equation}

\end{itemize}
At the tree-level, the gluon propagator stemming from the action \eqref{rgzlcg2} is 

\begin{equation}
\langle A^{a}_{\mu}(k)A^{b}_{\nu}(-k)\rangle = \delta^{ab}\left[ \frac{k^2 + M^2}{(k^2+m^2)(k^2+M^2) + 2g^2N \gamma^4} \left( \delta_{\mu\nu} - \frac{k_\mu k_\nu}{k^2} \right)+\frac{\alpha}{k^2}\frac{k_{\mu}k_{\nu}}{k^2}\right]\,.
\label{rgzlcg3}
\end{equation}
At this order, the transverse component of the propagator is $\alpha$-independent, but one should keep in mind that as long as higher loops are considered, $\alpha$-dependent corrections might appear. Though, we underline that the pole mass of the transverse component retains its independence from $\alpha$ to all orders \cite{Capri:2016aqq,Capri:2015nzw,Capri:2016gut}. For the longitudinal part, as previously mentioned, the result is exact. The tree-level propagator \eqref{rgzlcg3} is in good agreement with the most recent lattice data in the linear covariant gauges \cite{Bicudo:2015rma}. As in the Landau gauge, a massive/decoupling behavior is observed for the transverse component. \\\\Owing to the aforementioned prescription for a BRST-invariant (R)GZ construction in the linear covariant gauges, it was established in \cite{Capri:2015nzw} that, in great similarity with the Landau gauge, the formation of the refining condensates happens in $d=3,4$ but not in $d=2$. Also, matter fields were introduced in \cite{Capri:2017abz} according to the prescription developed in \cite{Capri:2014bsa}, giving rise to analytic expressions for the non-perturbative propagators for scalar fields in the adjoint representation of the gauge group as well as for quarks in the fundamental representation.  

%%%%%%%%%%%%%%%%%%%%%%%%%%%%%%%%%%%%%%%%%%%%%%%%%%%%%%%%%%%%%%%%
\section{Renormalizability analysis: Preliminaries} \label{Renpreli}
%%%%%%%%%%%%%%%%%%%%%%%%%%%%%%%%%%%%%%%%%%%%%%%%%%%%%%%%%%%%%%%%

%%%%%%%%%%%%%%%%%%%%%%%%%%%%%%%%%%%%%%%%%%%%%%%%%%%%%%%%%%%%%%%%
\subsection{Conventions}
%%%%%%%%%%%%%%%%%%%%%%%%%%%%%%%%%%%%%%%%%%%%%%%%%%%%%%%%%%%%%%%%

In order to give an algebraic proof of the renormalizability of the action \eqref{rgzlcg2}, it is convenient, in analogy with \cite{Fiorentini:2016rwx}, to adopt the following parametrization:

\begin{equation}
A^{a}_{\mu}\to  \frac{1}{g}\,A^{a}_{\mu}\,,\qquad
	b^{a}\to gb^{a}\,, \qquad \xi^a \to \frac{1}{g} \xi^a\;, \qquad \alpha\to\frac{\alpha}{g^{2}}\;, \qquad m^2 \to m^2 g^2\;, \qquad \tau^a\to g \tau^a \,, 
\label{ra1}   
\end{equation}
As a consequence, the Faddeev-Popov action in the linear covariant gauges is rewritten as

\begin{equation}
S^{\mathrm{FP}}_{\mathrm{LCG}} = \frac{1}{4g^2}\int \dd^4x~F^{a}_{\mu\nu}F^{a}_{\mu\nu} + \int \dd^4x\left(ib^a \partial_{\mu}A^{a}_{\mu}+ \frac{\alpha}{2}b^a b^a + \bar{c}^a\partial_{\mu}D^{ab}_{\mu}c^b\right)\,,
\label{ra2}
\end{equation}
with

\begin{equation}
F^{a}_{\mu\nu} = \partial_{\mu}A^{a}_{\nu}-\partial_{\nu}A^{a}_{\mu}+f^{abc}A^{b}_{\mu}A^{c}_{\nu}\,,\qquad D^{ab}_{\mu} = \delta^{ab}\partial_{\mu}-f^{abc}A^{c}_{\mu}\,,
\label{ra3}
\end{equation}
the redefined field strength and covariant derivative. Accordingly, the RGZ action in the linear covariant gauges becomes

\begin{eqnarray}
S^{\mathrm{LCG}}_{\mathrm{RGZ}}&=& S^{\mathrm{LCG}}_{\mathrm{FP}}-\int \dd^4x\left[\bar{\varphi}^{ac}_{\mu}\EuScript{M}^{ab}(A^h){\varphi}^{bc}_{\mu}-\bar{\omega}^{ac}_\mu\EuScript{M}^{ab}(A^h)\omega^{bc}_\mu+\gamma^{2}f^{abc}(A^h)^a_{\mu}(\varphi+\bar{\varphi})^{bc}_\mu\right]\nonumber\\
&+&\frac{m^2}{2}\int\dd^4x~(A^h)^a_{\mu}(A^h)^a_{\mu}-M^2\int \dd^4x\left(\bar{\varphi}^{ab}_{\mu}\varphi^{ab}_{\mu}-\bar{\omega}^{ab}_{\mu}\omega^{ab}_{\mu}\right)+\int \dd^4x~\tau^a \partial_\mu (A^h)^a_\mu - \int \dd^4x~\bar{\eta}^a \EuScript{M}^{ab}(A^h)\eta^b\,,
\label{ra4}
\end{eqnarray}
with 

\begin{equation}
A^{h}_{\mu} \equiv (A^{h})^a_{\mu}\,T^{a}=h^{\dagger}A_{\mu}h+i h^{\dagger}\partial_{\mu}h \;, \qquad 
h=e^{i\xi^{a}T^{a}} \,.
\label{ra5} 
\end{equation}
After the redefinition \eqref{ra1}, the BRST transformations read

\begin{align}
sA^{a}_{\mu}&=-D^{ab}_{\mu}c^b\,,     &&sc^a=\frac{1}{2}f^{abc}c^bc^c\,, \nonumber\\
s\bar{c}^a&=ib^{a}\,,     &&sb^{a}=0\,, \nonumber\\
s\varphi^{ab}_{\mu}&=0\,,   &&s\omega^{ab}_{\mu}=0\,, \nonumber\\
s\bar{\omega}^{ab}_{\mu}&=0\,,         &&s\bar{\varphi}^{ab}_{\mu}=0\,,\nonumber \\
s \xi^a & = g^{ab}(\xi) c^b  \,, && sA^{h,a}_\mu =0\,,   \nonumber \\
s\tau^a& =0\,, && s\bar{\eta}^a=0 \,, \nonumber \\
s\eta^a &= 0\,, && s^2 = 0\,,
\label{ra6}
\end{align}
with

\begin{equation}
g^{ab}(\xi) = - \delta^{ab} + \frac{1}{2} f^{abc} \xi^c - \frac{1}{12} f^{acd} f^{cbe} \xi^e \xi^d + O(\xi^3)\,.
\label{ra7}
\end{equation}
Before writing the Ward identities, we need to introduced  a suitable set of external sources which we describe in details in the next subsection.

%%%%%%%%%%%%%%%%%%%%%%%%%%%%%%%%%%%%%%%%%%%%%%%%%%%%%%%%%%%%%%%%
\subsection{Introduction of external sources}
%%%%%%%%%%%%%%%%%%%%%%%%%%%%%%%%%%%%%%%%%%%%%%%%%%%%%%%%%%%%%%%%

Let us begin by introducing the following set of sources $(M,V,N,U)^{ab}_{\mu\nu}$ and express the term which contains the Gribov parameter $\gamma$ in \eqref{rgzlcg2} as 

\begin{eqnarray}
\int \dd^4x~\gamma^2f^{abc}(A^{h})^a_{\mu}(\varphi+\bar{\varphi})^{bc}_{\mu} &\longrightarrow& \int \dd^4x \left(M^{ai}_{\mu}D^{ab}_{\mu}(A^h)\varphi^{bi}+V^{ai}_{\mu}D^{ab}_{\mu}(A^h)\bar{\varphi}^{bi}+N^{ai}_{\mu}D^{ab}_{\mu}(A^h)\omega^{bi}\right.\nonumber\\
&+&\left.U^{ai}_{\mu}D^{ab}_{\mu}(A^h)\bar{\omega}^{bi}-M^{ai}_\mu V^{ai}_\mu +N^{ai}_\mu U^{ai}_\mu \right)\,,
\label{ra8}
\end{eqnarray}
where we are employing the multi-index notation $i=(a,\mu)$ in the same way as done in \cite{Dudal:2008sp,Zwanziger:1992qr,Maggiore:1993wq}. We also emphasize that the presence of terms which are quadratic in  the sources is allowed by power counting. The local sources $(M,V,N,U)^{ab}_{\mu\nu}$ enlarge the original theory \eqref{rgzlcg2} which is recovered by demanding that they  attain a  suitable physical limit,  namely 

\begin{eqnarray}
M^{ab}_{\mu\nu}\Big|_{\mathrm{phys}}&=& V^{ab}_{\mu\nu}\Big|_{\mathrm{phys}}=\gamma^2\delta^{ab}\delta_{\mu\nu}\,,\nonumber\\
N^{ab}_{\mu\nu}\Big|_{\mathrm{phys}}&=& U^{ab}_{\mu\nu}\Big|_{\mathrm{phys}}=0\,,
\label{ra9}
\end{eqnarray}
from which the right-hand side of eq.\eqref{ra8} reduces precisely to the left-hand side.  
Also, in order to preserve the BRST invariance of the theory, the sources are chosen to be BRST-singlets, \textit{i.e.}

\begin{equation}
s M^{ai}_{\mu}=s V^{ai}_{\mu}= s N^{ai}_{\mu} = s U^{ai}_{\mu} = 0\,,
\label{ra10}
\end{equation}
In addition,  following the procedure of the algebraic renormalization setup \cite{Piguet:1984js}, additional external sources coupled to the composite operators corresponding to the non-linear BRST transformations of the fields need to be introduced, {\it i.e.}

\begin{equation}
S_{\mathrm{sources}} = \int \dd^4x\left[-\Omega^{a}_\mu D^{ab}_{\mu}c^b+\frac{1}{2}L^a f^{abc}c^{b}c^{c}+\mathcal{J}^{a}_{\mu}(A^{h})^a_{\mu}+K^{a}g^{ab}(\xi)c^b\right]\,,
\label{ra11}
\end{equation}
where the composite operator $(A^{h})^a_{\mu}$ has  also been coupled to its corresponding source $\mathcal{J}^{a}_{\mu}$,  see \cite{Fiorentini:2016rwx}. Finally, according to the local composite operator method (LCO)  \cite{Knecht:2001cc,Verschelde:2001ia} for  evaluating the effective potential giving rise to the dimension two  condensates $\langle A^{ha}_{\mu}A^{ha}_{\mu}\rangle$ and $\langle \bar{\omega}^{ai}\omega^{ai}-\bar{\varphi}^{ai}\varphi^{ai} \rangle$, two external sources $J$ and $\tilde{J}$ coupled to the corresponding  operators $(A^{ha}_{\mu}(x)A^{ha}_{\mu}(x))$ and  $(\bar{\omega}^{ai}(x)\omega^{ai}(x)-\bar{\varphi}^{ai}(x)\varphi^{ai}(x)) $, need  be  introduced,

\begin{equation}
S_{\mathrm{cond}}=\int \dd^4x\left[J (A^{h})^a_{\mu}(A^{h})^a_{\mu}+\tilde{J}(\bar{\omega}^{ai}\omega^{ai}-\bar{\varphi}^{ai}\varphi^{ai})+\frac{\theta}{2}J^2\right]\,,
\label{ra12}
\end{equation} 
where the parameter $\theta$ appearing in the quadratic source term $J^2$  of eq.\eqref{ra12} takes into account the UV divergences present in the vacuum correlation function $\langle ((A^{h})^a_{\mu})^2 (x) ((A^{h})^b_{\nu})^2 (y)\rangle$ when $x\to y$. Moreover, as argued in \cite{Dudal:2008sp}, it is not necessary to add a quadratic term in $\tilde{J}$, due to the absence of UV divergences in  $\langle  (\bar{\omega}^{ai} \omega^{ai}-\bar{\varphi}^{ai}\varphi^{ai})_x (\bar{\omega}^{ai}\omega^{ai}-\bar{\varphi}^{ai}\varphi^{ai})_y\rangle$, for ${x\to y}$. \\\\For future use, it will be also necessary to introduce an extra term depending on external sources given by
\begin{equation}
S_{\mathrm{extra}}=\int \dd^4x\left[\,-\Xi^{a}_{\mu}\,D^{ab}_{\mu}(A^{h})\eta^{b}
+X^{i}\,\eta^{a}\bar{\omega}^{ai}
+Y^{i}\,\eta^{a}\bar{\varphi}^{ai}
+\bar{X}^{abi}\,\eta^{a}\omega^{bi}+\bar{Y}^{abi}\,\eta^{a}\varphi^{bi}\,\right]\,.
\end{equation}
The whole new set of sources is invariant under BRST transformations, \textit{i.e.}

\begin{equation}
s\Omega^a_\mu = sL^a = s\mathcal{J}^a_\mu = sK^a = sJ = s\tilde{J}=
s\Xi^{a}_{\mu}=sX^{i}=sY^{i}=s\bar{X}^{i}=s\bar{Y}^{i}=0\,.
\label{ra13}
\end{equation}
After the introduction of the external sources, for the complete extended classical action $\Sigma$ one has 

\begin{eqnarray}
\Sigma &=& S^{\mathrm{LCG}}_{\mathrm{FP}}-\int \dd^4x\left(\bar{\varphi}^{ac}_{\mu}\EuScript{M}^{ab}(A^h){\varphi}^{bc}_{\mu}-\bar{\omega}^{ac}_\mu\EuScript{M}^{ab}(A^h)\omega^{bc}_\mu\right)+\int \dd^4x~\tau^a \partial_\mu (A^h)^a_\mu - \int \dd^4x~\bar{\eta}^a \EuScript{M}^{ab}(A^h)\eta^b\nonumber\\
&-&\int \dd^4x \left(M^{ai}_{\mu}D^{ab}_{\mu}(A^h)\varphi^{bi}+V^{ai}_{\mu}D^{ab}_{\mu}(A^h)\bar{\varphi}^{bi}-N^{ai}_{\mu}D^{ab}_{\mu}(A^h)\omega^{bi}+U^{ai}_{\mu}D^{ab}_{\mu}(A^h)\bar{\omega}^{bi}+M^{ai}_\mu V^{ai}_\mu -N^{ai}_\mu U^{ai}_\mu \right)\nonumber\\
&+& \int \dd^4x\left[-\Omega^{a}_\mu D^{ab}_{\mu}c^b+\frac{1}{2}L^a f^{abc}c^{b}c^{c}+\mathcal{J}^{a}_{\mu}(A^{h})^a_{\mu}+K^{a}g^{ab}(\xi)c^b\right] + \int \dd^4x\left[J (A^{h})^a_{\mu}(A^{h})^a_{\mu}+\tilde{J}(\bar{\omega}^{ai}\omega^{ai}-\bar{\varphi}^{ai}\varphi^{ai})+\frac{\theta}{2}J^2\right]\nonumber\\
&+&\int \dd^4x\left[\,-\Xi^{a}_{\mu}\,D^{ab}_{\mu}(A^{h})\eta^{b}
+X^{i}\,\eta^{a}\bar{\omega}^{ai}
+Y^{i}\,\eta^{a}\bar{\varphi}^{ai}
+\bar{X}^{abi}\,\eta^{a}\omega^{bi}
+\bar{Y}^{abi}\,\eta^{a}\varphi^{bi}\,\right]\,,
\label{ra14}
\end{eqnarray}
with 
\begin{equation} 
s \Sigma = 0 \;. \label{sS} 
\end{equation}

%%%%%%%%%%%%%%%%%%%%%%%%%%%%%%%%%%%%%%%%%%%%%%%%%%%%%%%%%%%%
\subsection{Extended BRST symmetry}
%%%%%%%%%%%%%%%%%%%%%%%%%%%%%%%%%%%%%%%%%%%%%%%%%%%%%%%%%%%%

As dicussed in \cite{Capri:2016gut,Piguet:1984js}, it turns out to be convenient to extend the action of the BRST operator $s$ on the parameter $\alpha$ as  

\begin{equation}
s\alpha = \chi\,,
\label{ra15}
\end{equation}
where $\chi$ is a constant Grassmann parameter with ghost number 1, to be set to zero at the end of the algebraic analysis. Following \cite{Piguet:1984js},  such a transformation plays a pivotal role in order to control the gauge parameter (in)dependence of correlation functions. For the renormalizability proof we shall present, we employ \eqref{ra15} as well. Furthermore, it can be shown that, besides the BRST invariance, eq.\eqref{sS}, the action \eqref{ra14} enjoys a second nilpotent exact symmetry:

\begin{equation} 
\delta \Sigma = 0\;, \label{deltaS}
\end{equation}
where $\delta$ is given by 
\begin{eqnarray}
\delta\varphi^{ai}&=&\omega^{ai}\,,\qquad \delta\omega^{ai}=0\nonumber\\
\delta\bar{\omega}^{ai}&=&\bar{\varphi}^{ai}\,,\qquad \delta\bar{\varphi}^{ai}=0\nonumber\\
\delta N^{ai}_\mu &=& M^{ai}_\mu\,,\qquad \delta M^{ai}_\mu=0\nonumber\\
\delta V^{ai}_\mu &=&U^{ai}_\mu\,,\qquad \delta U^{ai}_\mu=0 \,,\nonumber\\
\delta Y^{i}&=&X^{i}\,,\qquad \delta X^{i}=0 \,,\nonumber\\
\delta \bar{X}^{abi}&=&-\bar{Y}^{abi}\,,\qquad \delta \bar{Y}^{abi}=0 \,, 
\label{ra16}
\end{eqnarray}
with $\delta^2 = 0$. The transformations \eqref{ra16} reveal a doublet structure for the localizing Zwanziger fields and sources.  Moreover, by taking into account that $\left\{s,\delta\right\}=0$, one can define a single extended nilpotent operator $\EuScript{Q}$ defined by

\begin{equation}
\EuScript{Q} = s + \delta\,, \qquad \EuScript{Q}^2 = 0\,.
\label{ra17}
\end{equation}
Also, for future applications, it turns out to be useful to embed the source $\tilde{J}$ into a  $\EuScript{Q}$-doublet by means of the introduction of the source $H$, transforming as  

\begin{equation}
\EuScript{Q}\tilde{J} = H\,,\qquad \EuScript{Q}H = 0\,.
\label{ra18}
\end{equation}
Summarizing, the full set of extended $\EuScript{Q}$-transformations are given by

\begin{align}
\EuScript{Q}A^{a}_{\mu}&=-D^{ab}_{\mu}c^b\,,     &&\EuScript{Q}c^a=\frac{1}{2}f^{abc}c^bc^c\,, \nonumber\\
\EuScript{Q}\bar{c}^a&=ib^{a}\,,     &&\EuScript{Q}b^{a}=0\,, \nonumber\\
\EuScript{Q}\varphi^{ab}_{\mu}&=\omega^{ab}_\mu \,,   &&\EuScript{Q}\omega^{ab}_{\mu}=0\,, \nonumber\\
\EuScript{Q}\bar{\omega}^{ab}_{\mu}&=\bar{\varphi}^{ab}_\mu\,,         &&\EuScript{Q}\bar{\varphi}^{ab}_{\mu}=0\,,\nonumber \\
\EuScript{Q}\xi^a & = g^{ab}(\xi) c^b  \,, && \EuScript{Q}A^{h,a}_\mu =0\,,   \nonumber \\
\EuScript{Q}\tau^a& =0\,, && \EuScript{Q}\bar{\eta}^a=0 \,, \nonumber \\
\EuScript{Q}\eta^a &= 0\,, && \EuScript{Q}\alpha = \chi\,, \nonumber \\
\EuScript{Q}\chi &= 0\,, && \EuScript{Q}N^{ai}_\mu = M^{ai}_\mu\,, \nonumber \\
\EuScript{Q} V^{ai}_\mu & = U^{ai}_\mu\,, && \EuScript{Q} U^{ai}_\mu=0\,, \nonumber \\
\EuScript{Q} Y^{i}& = X^{i}\,, && \EuScript{Q} X^{i}=0\,,\nonumber\\
\EuScript{Q} \bar{X}^{abi}& =-\bar{Y}^{abi}\,, && \EuScript{Q} \bar{Y}^{abi}=0\,,\nonumber\\
\EuScript{Q}\Omega^a_\mu & = 0\,, && \EuScript{Q}L^a = 0\,, \nonumber\\
\EuScript{Q}K^a & = 0\,, && \EuScript{Q}\mathcal{J}^{a}_\mu = 0\,, \nonumber\\
\EuScript{Q}J & = 0\,, && \EuScript{Q}\tilde{J} = H\,, \nonumber\\
\EuScript{Q}H & = 0\,, && \EuScript{Q}\Xi^{a}_{\mu} = 0\,.
\label{ra19}
\end{align}
%$\spadesuit$ and 
%\begin{equation}
%\EuScript{Q}^{2}=0\,.
%\end{equation}
The complete classical action $\Sigma$ invariant under the extended transformations \eqref{ra19} is, explicitly,

\begin{eqnarray}
\Sigma &=& \frac{1}{4g^2}\int \dd^4x~F^{a}_{\mu\nu}F^{a}_{\mu\nu}+\int \dd^4x\left(ib^a \partial_\mu A^a_\mu+\frac{\alpha}{2}b^a b^a-\frac{i}{2}\chi \bar{c}^a b^a + \bar{c}^a\partial_\mu D^{ab}_\mu c^b\right)-\int \dd^4x\left(\bar{\varphi}^{ac}_{\mu}\EuScript{M}^{ab}(A^h){\varphi}^{bc}_{\mu}\right.\nonumber\\
&-&\left.\bar{\omega}^{ac}_\mu\EuScript{M}^{ab}(A^h)\omega^{bc}_\mu\right)+\int \dd^4x~\tau^a \partial_\mu (A^h)^a_\mu - \int \dd^4x~\bar{\eta}^a \EuScript{M}^{ab}(A^h)\eta^b-\int \dd^4x \left(M^{ai}_{\mu}D^{ab}_{\mu}(A^h)\varphi^{bi}+V^{ai}_{\mu}D^{ab}_{\mu}(A^h)\bar{\varphi}^{bi}\right.\nonumber\\
&-&\left.N^{ai}_{\mu}D^{ab}_{\mu}(A^h)\omega^{bi}+U^{ai}_{\mu}D^{ab}_{\mu}(A^h)\bar{\omega}^{bi}+M^{ai}_\mu V^{ai}_\mu -N^{ai}_\mu U^{ai}_\mu \right)+ \int \dd^4x\left[-\Omega^{a}_\mu D^{ab}_{\mu}c^b+\frac{1}{2}L^a f^{abc}c^{b}c^{c}+\mathcal{J}^{a}_{\mu}(A^{h})^a_{\mu}\right.\nonumber\\
&+&\left.K^{a}g^{ab}(\xi)c^b\right] + \int \dd^4x\left[J (A^{h})^a_{\mu}(A^{h})^a_{\mu}+\tilde{J}(\bar{\omega}^{ai}\omega^{ai}-\bar{\varphi}^{ai}\varphi^{ai})+H\bar{\omega}^{ai}\varphi^{ai}+\frac{\theta}{2}J^2\right]\nonumber\\
&+&\int \dd^4x\left[\,-\Xi^{a}_{\mu}\,D^{ab}_{\mu}(A^{h})\eta^{b}
+X^{i}\,\eta^{a}\bar{\omega}^{ai}
+Y^{i}\,\eta^{a}\bar{\varphi}^{ai}
+\bar{X}^{abi}\,\eta^{a}\omega^{bi}
+\bar{Y}^{abi}\,\eta^{a}\varphi^{bi}\,\right]\,,
\label{ra20}
\end{eqnarray}
with 
\begin{equation} 
\EuScript{Q} \Sigma = 0 \;. \label{qS}
\end{equation}
As already pointed out, the action $\Sigma$  is an enlarged action which reduces to the original one \eqref{ra4} when the sources attain suitable physical values, summarized below 

\begin{eqnarray}
\chi\Big|_{\mathrm{phys}} &=&  \Omega^a_\mu\Big|_{\mathrm{phys}} = L^a\Big|_{\mathrm{phys}} = K^a\Big|_{\mathrm{phys}} = \mathcal{J}^a_\mu\Big|_{\mathrm{phys}} = H\Big|_{\mathrm{phys}} = 0\,, \nonumber \\
\Xi^{a}_{\mu}\Big|_{\mathrm{phys}} &=& X^{i}\Big|_{\mathrm{phys}} = Y^{i}\Big|_{\mathrm{phys}} = \bar{X}^{abi}\Big|_{\mathrm{phys}} = \bar{Y}^{abi}\Big|_{\mathrm{phys}} =0\,,\nonumber\\
\qquad J\Big|_{\mathrm{phys}} &=&  \frac{m^2}{2}\,,\qquad \tilde{J}\Big|_{\mathrm{phys}}=M^2\,,  \nonumber \\
M^{ab}_{\mu\nu}\Big|_{\mathrm{phys}}&=& V^{ab}_{\mu\nu}\Big|_{\mathrm{phys}}=\gamma^2\delta^{ab}\delta_{\mu\nu}\,, \qquad 
N^{ab}_{\mu\nu}\Big|_{\mathrm{phys}}= U^{ab}_{\mu\nu}\Big|_{\mathrm{phys}}=0\,,
\label{ra21}
\end{eqnarray}
so that 
\begin{equation} 
\Sigma\Big|_{\mathrm{phys}} =  S^{\mathrm{LCG}}_{\mathrm{RGZ}} \;. \label{pl}
\end{equation}
Since the action $S^{\mathrm{LCG}}_{\mathrm{RGZ}}$ can be seen as a particular case of the more general extended action $\Sigma$,  the renormalizability of $\Sigma$ will imply that of $S^{\mathrm{LCG}}_{\mathrm{RGZ}}$. \\\\Let us also notice that,  thanks to the parametrization \eqref{ra1}, it is simple to check that

\begin{equation}
g^2\frac{\partial\Sigma}{\partial g^2} = -\frac{1}{4g^2}\int \dd^4x~F^{a}_{\mu\nu}F^{a}_{\mu\nu}\,,
\label{ra22}
\end{equation}
a feature that will be exploited in the proof of the renormalizability. For the benefit of the reader, the quantum numbers of fields, sources and parameters of the theory are collected in tables \ref{table1} and \ref{table2}. 

\begin{table}
\centering
\begin{tabular}{|c|c|c|c|c|c|c|c|c|c|c|c|c|c|c|}
\hline
Fields &$A$&$b$&$c$&$\bar{c}$&$\xi$&$\bar{\varphi}$&$\varphi$&$\bar{\omega}$&$\omega$&$\alpha$&$\chi$&$\tau$&$\eta$&$\bar{\eta}$\\
\hline\hline
Dimension &1&2&0&2&0&1&1&1&1&0&0&2&0&2\\
\hline
$c$-ghost number&0&0&1&$-1$&0&0&0&$-1$&1&0&1&0&0&0\\
\hline
$\eta$-ghost number&0&0&0&0&0&0&0&0&0&0&0&0&1&$-1$\\
\hline
$U(f)$-charge &0&0&0&0&0&$-1$&1&$-1$&1&0&0&0&0&0\\
\hline
Nature &B&B&F&F&B&B&B&F&F&B&F&B&F&F\\
\hline
\end{tabular}
\caption{The quantum numbers of fields.}
\label{table1}
\end{table}

\begin{table}
\centering
\begin{tabular}{|c|c|c|c|c|c|c|c|c|c|c|c|c|c|c|c|c|}
\hline
Sources &$\Omega$&$L$&$K$&$J$&$\mathcal{J}$&$M$&$N$&$U$&$V$&$\tilde{J}$&$H$&$\Xi$&$X$&$Y$&$\bar{X}$&$\bar{Y}$\\
\hline\hline
Dimension &3&4&4&2&3&2&2&2&2&2&2&3&3&3&3&3\\
\hline
$c$-ghost number&$-1$&$-2$&$-1$&0&0&0&$-1$&$1$&0&0&1&0&1&0&$-1$&0\\
\hline
$\eta$-ghost number&0&0&0&0&0&0&0&0&0&0&0&$-1$&$-1$&$-1$&$-1$&$-1$\\
\hline
$U(f)$-charge &0&0&0&0&0&$-1$&$-1$&1&1&0&0&0&1&1&$-1$&$-1$\\
\hline
Nature &F&B&F&B&B&B&F&F&B&B&F&F&B&F&B&F\\
\hline
\end{tabular}
\caption{The quantum numbers of sources.}
\label{table2}
\end{table}

%%%%%%%%%%%%%%%%%%%%%%%%%%%%%%%%%%%%%%%%%%%%%%%%%%%%%%%%%%%%%%
\section{Ward identities and algebraic characterization of the most general counterterm} \label{Wardid}
%%%%%%%%%%%%%%%%%%%%%%%%%%%%%%%%%%%%%%%%%%%%%%%%%%%%%%%%%%%%%%

The classical extended action $\Sigma$ defined by eq.\eqref{ra20} enjoys a rich set of symmetries characterized by the following Ward identities,

\begin{itemize}

\item Slavnov-Taylor identity 

\begin{equation}
\mathcal{S}_{\EuScript{Q}}(\Sigma)= 0\,,
\label{wi0}
\end{equation}
with

\begin{eqnarray}
\mathcal{S}_{\EuScript{Q}}(\Sigma)&=&\int \dd^4x \left(\frac{\delta\Sigma}{\delta A^{a}_{\mu}}\frac{\delta\Sigma}{\delta\Omega^{a}_{\mu}}+\frac{\delta\Sigma}{\delta c^a}\frac{\delta\Sigma}{\delta L^a}+\frac{\delta\Sigma}{\delta\xi^a}\frac{\delta\Sigma}{\delta K^a}
+ib^a\frac{\delta\Sigma}{\delta\bar{c}^a}
+\omega^{ai}\frac{\delta\Sigma}{\delta\varphi^{ai}}
+\bar{\varphi}^{ai}\frac{\delta\Sigma}{\delta\bar{\omega}^{ai}}
+M^{ai}_\mu\frac{\delta\Sigma}{\delta N^{ai}_\mu}
\right.\nonumber\\
&+&\left.U^{ai}_\mu\frac{\delta\Sigma}{\delta V^{ai}_\mu}
+H\frac{\delta\Sigma}{\delta \tilde{J}}
+X^{i}\frac{\delta\Sigma}{\delta Y^{i}}
-\bar{Y}^{abi}\frac{\delta\Sigma}{\delta \bar{X}^{abi}}
\right)+\chi\frac{\partial\Sigma}{\partial\alpha}\,.
\label{wi1}
\end{eqnarray}

\item Anti-ghost equation

\begin{equation}
\frac{\delta\Sigma}{\delta \bar{c}^a}+\partial_{\mu}\frac{\delta \Sigma}{\delta \Omega^a_{\mu}}=\frac{i}{2}\chi b^a\,.
\label{wi2}
\end{equation}

\item Gauge-fixing condition

\begin{equation}
\frac{\delta\Sigma}{\delta b^a} = i\partial_\mu A^a_\mu + \alpha b^a-\frac{i}{2}\chi\bar{c}^a\,.
\label{wi3}
\end{equation}

\item Equation of motion of the Lagrange multiplier $\tau^a$

\begin{equation}
\frac{\delta\Sigma}{\delta \tau^a}-\partial_{\mu}\frac{\delta\Sigma}{\delta \mathcal{J}^{a}_{\mu}}=0\,.
\label{wi4}
\end{equation}

\item Global $U(f)$ symmetry

\begin{equation}
U_{ij}\Sigma = 0\,,
\label{wi5}
\end{equation}
with
\begin{eqnarray}
U_{ij}&=&\int \dd^4x\left(\varphi^{ai}\frac{\delta}{\delta\varphi^{aj}}-\bar{\varphi}^{aj}\frac{\delta}{\delta\bar{\varphi}^{ai}}+\omega^{ai}\frac{\delta}{\delta\omega^{aj}}-\bar{\omega}^{aj}\frac{\delta}{\delta\bar{\omega}^{ai}}-M^{aj}_\mu\frac{\delta}{\delta M^{ai}_\mu}+V^{ai}_\mu\frac{\delta}{\delta V^{aj}_\mu}\right.\nonumber\\
&-&\left.N^{aj}\frac{\delta}{\delta N^{ai}}+U^{ai}\frac{\delta}{\delta U^{aj}}
+X^{i}\frac{\delta}{\delta X^{j}}
+Y^{i}\frac{\delta}{\delta Y^{j}}
-\bar{X}^{abj}\frac{\delta}{\delta \bar{X}^{abi}}
-\bar{Y}^{abj}\frac{\delta}{\delta \bar{Y}^{abi}}\right)\,.
\label{wi6}
\end{eqnarray}

\item Linearly broken constraints

\begin{equation}
\frac{\delta\Sigma}{\delta\bar{\varphi}^{ai}}+\partial_\mu\frac{\delta\Sigma}{\delta M^{ai}_\mu}+f^{abc}V^{bi}_{\mu}\frac{\delta\Sigma}{\delta\mathcal{J}^{c}_{\mu}}=-\tilde{J}\varphi^{ai}+Y^{i}\eta^{a}\,,
\label{wi7}
\end{equation}

\begin{equation}
\frac{\delta\Sigma}{\delta\varphi^{ai}}
+\partial_{\mu}\frac{\delta\Sigma}{\delta V^{ai}_{\mu}}
-f^{abc}\bar{\varphi}^{bi}\frac{\delta\Sigma}{\delta\tau^{c}}
+f^{abc}M^{bi}_{\mu}\frac{\delta\Sigma}{\delta\mathcal{J}^{c}_{\mu}}=-\tilde{J}\bar{\varphi}^{ai}
-H\bar{\omega}^{ai}+\bar{Y}^{bai}\eta^{b}\,,
\label{wi8}
\end{equation}

\begin{equation}
\frac{\delta\Sigma}{\delta\bar{\omega}^{ai}}+\partial_\mu\frac{\delta\Sigma}{\delta N^{ai}_\mu}-f^{abc}U^{bi}_{\mu}\frac{\delta\Sigma}{\delta\mathcal{J}^{c}_{\mu}}=\tilde{J}\omega^{ai}-H\varphi^{ai}-X^{i}\eta^{a}\,,
\label{wi9}
\end{equation}

\begin{equation}
\frac{\delta\Sigma}{\delta\omega^{ai}}
+\partial_{\mu}\frac{\delta\Sigma}{\delta U^{ai}_{\mu}}
-f^{abc}\bar{\omega}^{bi}\frac{\delta\Sigma}{\delta\tau^{c}}
+f^{abc}N^{bi}_{\mu}\frac{\delta\Sigma}{\delta\mathcal{J}^{c}_{\mu}}=-\tilde{J}\bar{\omega}^{ai}
-\bar{X}^{bai}\eta^{a}\,.
\label{wi10}
\end{equation}

\item $c$-ghost number and $\eta$-ghost number Ward identities

\begin{eqnarray}
&\phantom{+}&\int \dd^4x\left(c^a\frac{\delta\Sigma}{\delta c^a}-\bar{c}^a\frac{\delta\Sigma}{\delta\bar{c}^a}+\omega^{ai}\frac{\delta\Sigma}{\delta\omega^{ai}}-\bar{\omega}^{ai}\frac{\delta\Sigma}{\delta\bar{\omega}^{ai}}-\Omega^{a}_{\mu}\frac{\delta\Sigma}{\delta\Omega^{a}_{\mu}}-2L^a\frac{\delta\Sigma}{\delta L^a}-K^a\frac{\delta\Sigma}{\delta K^a}+U^{ai}\frac{\delta\Sigma}{\delta U^{ai}}\right.\nonumber\\
&&-\left.N^{ai}_{\mu}\frac{\delta\Sigma}{\delta N^{ai}_{\mu}}-\tilde{J}\frac{\partial\Sigma}{\delta\tilde{J}}
+X^{i}\frac{\delta\Sigma}{\delta X^{i}}
-\bar{X}^{abi}\frac{\delta\Sigma}{\delta\bar{X}^{abi}} \right) 
+\chi\frac{\partial\Sigma}{\partial\chi}=0\,,
\label{wi11}
\end{eqnarray}
\begin{eqnarray}
&& \int \dd^4x\left(\eta^{a}\frac{\delta\Sigma}{\delta \eta^{a}}
-\bar{\eta}^{a}\frac{\delta\Sigma}{\delta \bar{\eta}^{a}}
-\Xi^{a}_{\mu}\frac{\delta\Sigma}{\delta \Xi^{a}_{\mu}}
-X^{i}\frac{\delta\Sigma}{\delta X^{i}}
-Y^{i}\frac{\delta\Sigma}{\delta Y^{i}}
-\bar{X}^{abi}\frac{\delta\Sigma}{\delta \bar{X}^{abi}}
-\bar{Y}^{abi}\frac{\delta\Sigma}{\delta \bar{Y}^{abi}}\right)=0\,.
\end{eqnarray}
\item Exact $\mathcal{R}_{ij}$ symmetry

\begin{equation}
\mathcal{R}_{ij}\Sigma=0\,,
\label{wi13}
\end{equation}
with

\begin{equation}
\mathcal{R}_{ij}=\int \dd^4x\left(\varphi^{ai}\frac{\delta}{\delta {\omega}^{aj}}-\bar{\omega}^{aj}\frac{\delta}{\delta \bar{\varphi}^{ai}}+V^{ai}_{\mu}\frac{\delta}{\delta U^{aj}_{\mu}}-N^{aj}_{\mu}\frac{\delta}{\delta M^{ai}_\mu}
+\bar{X}^{abj}\frac{\delta}{\delta\bar{Y}^{abi}}+Y^{i}\frac{\delta}{\delta X^{j}}\right)\,.
\label{wi14}
\end{equation}

\item Local $\bar{\eta}$ equation

\begin{equation}
\frac{\delta \Sigma}{\delta \bar{\eta}^a}+\partial_{\mu}\frac{\delta\Sigma}{\delta\Xi^{a}_{\mu}}=0\,. 
\label{w15}
\end{equation}

\item Integrated linearly broken $\eta$ equation

\begin{equation}
\int \dd^4x\left(\frac{\delta\Sigma}{\delta \eta^a}+f^{abc}\bar{\eta}^b\frac{\delta \Sigma}{\delta \tau^c}
-f^{abc}\Xi^{b}_{\mu}\frac{\delta\Sigma}{\delta\mathcal{J}^{c}_{\mu}}\right)=
\int \dd^4x\left(-\bar{Y}^{abi}\varphi^{bi}+\bar{X}^{abi}\omega^{bi}+X\bar{\omega}^{ai}-Y^{i}\bar{\varphi}^{ai}\right)\,.
\label{w16}
\end{equation}

\item Identities that mix the Zwanziger ghosts with the new ghosts
\begin{eqnarray}
W^{i}_{(1)}(\Sigma)&=&\int \dd^4x \left(
\bar{\omega}^{ai}\frac{\delta\Sigma}{\delta\bar{\eta}^{a}}
+\eta^{a}\frac{\delta\Sigma}{\delta\omega^{ai}}
+N^{ai}_{\mu}\frac{\delta\Sigma}{\delta\Xi^{a}_{\mu}}
+\tilde{J}\frac{\delta\Sigma}{\delta X^{i}}
\right)=0\,,\\
W^{i}_{(2)}(\Sigma)&=&\int \dd^4x \left(
\bar{\varphi}^{ai}\frac{\delta\Sigma}{\delta\bar{\eta}^{a}}
-\eta^{a}\frac{\delta\Sigma}{\delta\varphi^{ai}}
+M^{ai}_{\mu}\frac{\delta\Sigma}{\delta\Xi^{a}_{\mu}}
-\tilde{J}\frac{\delta\Sigma}{\delta Y^{i}}
+H\frac{\delta\Sigma}{\delta X^{i}}
\right)=0\,,\\
W^{i}_{(3)}(\Sigma)&=&\int \dd^4x \left(
\varphi^{ai}\frac{\delta\Sigma}{\delta\bar{\eta}^{a}}
-\eta^{a}\frac{\delta\Sigma}{\delta\bar{\varphi}^{ai}}
-f^{abc}\frac{\delta\Sigma}{\delta\bar{Y}^{abi}}\frac{\delta\Sigma}{\delta\tau^{c}}
-V^{ai}_{\mu}\frac{\delta\Sigma}{\delta\Xi^{a}_{\mu}}
+\tilde{J}\frac{\delta\Sigma}{\delta\bar{Y}^{aai}}
\right)=0\,,\\
W^{i}_{(4)}(\Sigma)&=&\int \dd^4x \left(
\omega^{ai}\frac{\delta\Sigma}{\delta\bar{\eta}^{a}}
-\eta^{a}\frac{\delta\Sigma}{\delta\bar{\omega}^{ai}}
+f^{abc}\frac{\delta\Sigma}{\delta\bar{X}^{abi}}\frac{\delta\Sigma}{\delta\tau^{c}}
+U^{ai}_{\mu}\frac{\delta\Sigma}{\delta\Xi^{a}_{\mu}}
+\tilde{J}\frac{\delta\Sigma}{\delta\bar{X}^{aai}}
+H\frac{\delta\Sigma}{\delta\bar{Y}^{aai}}
\right)=0\,.
\end{eqnarray}
\end{itemize}
In order to characterize the most general invariant counterterm, which can be freely added to all orders
in perturbation theory, we follow the setup of the algebraic renormalization  \cite{Piguet:1995er} and 
perturb the classical action $\Sigma$ by adding an integrated local quantity in the fields and sources, 
$\Sigma_{\mathrm{CT}}$, with dimension bounded by four and vanishing $c$ and $\eta$-ghost number. We demand thus that the 
perturbed action, $(\Sigma +\epsilon  \Sigma_{\mathrm{CT}})$, where $\epsilon$ is an expansion parameter, 
fulfills, to the first order in $\epsilon$, the same Ward identities obeyed by the classical action,  \textit{i.e.}

\begin{eqnarray}
&&\mathcal{S}_{\EuScript{Q}}(\Sigma + \epsilon \Sigma_{\mathrm{CT}})=\mathcal{O}(\epsilon^2)\,, \nonumber\\
&&\left(\frac{\delta}{\delta \bar{c}^a}+\partial_{\mu}\frac{\delta }{\delta \Omega^a_{\mu}}\right)(\Sigma+\epsilon \Sigma_{\mathrm{CT}})-\frac{i}{2}\chi b^a = \mathcal{O}(\epsilon^2)\,,\nonumber\\
&&\frac{\delta}{\delta b^a}(\Sigma+\epsilon \Sigma_{\mathrm{CT}}) = i\partial_\mu A^a_\mu + \alpha b^a-\frac{i}{2}\chi\bar{c}^a+\mathcal{O}(\epsilon^2)\,,\nonumber\\
&&\left(\frac{\delta}{\delta \tau^a}-\partial_{\mu}\frac{\delta}{\delta \mathcal{J}^{a}_{\mu}}\right)(\Sigma+\epsilon \Sigma_{\mathrm{CT}})=\mathcal{O}(\epsilon^2)\,,\nonumber\\
&& U_{ij}(\Sigma+\epsilon \Sigma_{\mathrm{CT}}) = \mathcal{O}(\epsilon^2)\,,\nonumber\\
&&\left(\frac{\delta}{\delta\bar{\varphi}^{ai}}+\partial_\mu\frac{\delta}{\delta M^{ai}_\mu}+f^{abc}V^{bi}_{\mu}\frac{\delta}{\delta\mathcal{J}^{c}_{\mu}}\right)(\Sigma+\epsilon \Sigma_{\mathrm{CT}})=-\tilde{J}\varphi^{ai}+Y^{i}\eta^{a}+\mathcal{O}(\epsilon^2)\,, \nonumber\\
&&\left(\frac{\delta}{\delta\varphi^{ai}}+\partial_{\mu}\frac{\delta}{\delta V^{ai}_{\mu}}-f^{abc}\bar{\varphi}^{bi}\frac{\delta}{\delta\tau^{c}}+f^{abc}M^{bi}_{\mu}\frac{\delta}{\delta\mathcal{J}^{c}_{\mu}}\right)(\Sigma+\epsilon \Sigma_{\mathrm{CT}})=-\tilde{J}\bar{\varphi}^{ai}
-H\bar{\omega}^{ai}+ \bar{Y}^{bai}\eta^{b}+\mathcal{O}(\epsilon^2)\nonumber\\
&&\left(\frac{\delta}{\delta\bar{\omega}^{ai}}+\partial_\mu\frac{\delta}{\delta N^{ai}_\mu}-f^{abc}U^{bi}_{\mu}\frac{\delta}{\delta\mathcal{J}^{c}_{\mu}}\right)(\Sigma+\epsilon \Sigma_{\mathrm{CT}})=\tilde{J}\omega^{ai}-H\varphi^{ai}
-X^{i}\eta^{a}+\mathcal{O}(\epsilon^2)\,,\nonumber\\
&&\left(\frac{\delta}{\delta\omega^{ai}}+\partial_{\mu}\frac{\delta}{\delta U^{ai}_{\mu}}-f^{abc}\bar{\omega}^{bi}\frac{\delta}{\delta\tau^{c}}+f^{abc}N^{bi}_{\mu}\frac{\delta}{\delta\mathcal{J}^{c}_{\mu}}\right)(\Sigma+\epsilon \Sigma_{\mathrm{CT}})=-\tilde{J}\bar{\omega}^{ai}
-\bar{X}^{bai}\eta^{b}+\mathcal{O}(\epsilon^2)\,,\nonumber\\
&& \mathcal{R}_{ij}(\Sigma+\epsilon \Sigma_{\mathrm{CT}})=\mathcal{O}(\epsilon^2)\,,\nonumber\\
&& \left( \frac{\delta}{\delta \bar{\eta}^a}+\partial_{\mu}\frac{\delta}{\delta\Xi^{a}_{\mu}}\right)(\Sigma+\epsilon \Sigma_{\mathrm{CT}})=\mathcal{O}(\epsilon^2)\,,\nonumber\\
&&\int \dd^4x\left(\frac{\delta}{\delta \eta^a}+f^{abc}\bar{\eta}^b\frac{\delta}{\delta \tau^c}
-f^{abc}\Xi^{b}_{\mu}\frac{\delta}{\delta\mathcal{J}^{c}_{\mu}}\right)(\Sigma+\epsilon \Sigma_{\mathrm{CT}})=
\int \dd^4x\left(-\bar{Y}^{abi}\varphi^{bi}+\bar{X}^{abi}\omega^{bi}+X\bar{\omega}^{ai}-Y^{i}\bar{\varphi}^{ai}\right)+\mathcal{O}(\epsilon^2)\,,\nonumber\\
&&W^{i}_{(1,2,3,4)}(\Sigma+\epsilon \Sigma_{\mathrm{CT}})=\mathcal{O}(\epsilon^2)\,.
\label{w17}
\end{eqnarray}
As a consequence of the first condition of eqs.\eqref{w17}, {\it i.e.} $\mathcal{S}_{\EuScript{Q}}(\Sigma + \epsilon \Sigma_{\mathrm{CT}})=\mathcal{O}(\epsilon^2)$, one gets  

\begin{equation}
\mathcal{B}_{\EuScript{Q}}\Sigma_{\mathrm{CT}} = 0\,,
\label{w18}
\end{equation}
where $\mathcal{B}_{\EuScript{Q}}$ is the so-called linearized nilpotent Slavnov-Taylor operator,

\begin{eqnarray}
\mathcal{B}_{\EuScript{Q}} &=& \int \dd^4x \left(
\frac{\delta\Sigma}{\delta A^{a}_{\mu}}\frac{\delta}{\delta\Omega^{a}_{\mu}}
+\frac{\delta\Sigma}{\delta\Omega^{a}_{\mu}}\frac{\delta}{\delta A^{a}_{\mu}}
+\frac{\delta\Sigma}{\delta c^a}\frac{\delta}{\delta L^a}
+\frac{\delta\Sigma}{\delta L^a}\frac{\delta}{\delta c^a}
+\frac{\delta\Sigma}{\delta\xi^a}\frac{\delta}{\delta K^a}
+\frac{\delta\Sigma}{\delta K^a}\frac{\delta}{\delta\xi^a}
+ib^a\frac{\delta}{\delta\bar{c}^a}
+\omega^{ai}\frac{\delta}{\delta\varphi^{ai}}\right.\nonumber\\
&+&\left.\bar{\varphi}^{ai}\frac{\delta}{\delta\bar{\omega}^{ai}}
+M^{ai}_\mu\frac{\delta}{\delta N^{ai}_\mu}
+U^{ai}_\mu\frac{\delta}{\delta V^{ai}_\mu}
+H\frac{\delta}{\delta\tilde{J}}
+X^{i}\frac{\delta}{\delta Y^{i}}
-\bar{Y}^{abi}\frac{\delta}{\delta \bar{X}^{abi}}
\right)+\chi\frac{\partial}{\partial\alpha}\,, 
\label{w19}
\end{eqnarray}
with 
\begin{equation}
\mathcal{B}_{\EuScript{Q}} \mathcal{B}_{\EuScript{Q}} =0 \;. \label{nln}
\end{equation}
Equation \eqref{w18} tells us that the invariant counterterm $\Sigma_{\mathrm{CT}}$ belongs to the cohomolgy of $\mathcal{B}_{\EuScript{Q}}$ in the space of the integrated local polynomials in the fields and sources with $c$ and $\eta$-ghost number zero and bounded by dimension four. Owing to the general results on the cohomology of Yang-Mills theories, see  \cite{Piguet:1995er},  the general solution of \eqref{w18} can be written as 

\begin{equation}
\Sigma_{\mathrm{CT}}=\Delta + \mathcal{B}_{\EuScript{Q}}\Delta^{(-1)}\,,
\label{w20}
\end{equation}  
with $\Delta$ and $\Delta^{(-1)}$ integrated local polynomials in the  fields and sources of dimension bounded by four and ghost number zero and minus one, respectively, and $\mathcal{B}_{\EuScript{Q}} \Delta = 0$, with $\Delta \neq \mathcal{B}_{\EuScript{Q}}(\ldots)$. At this point one sees the usefulness of the extended operator $\EuScript{Q}$. The auxiliary fields and sources introduced due to the  restriction of the functional measure to the Gribov region are doublets with respect to $\EuScript{Q}$, implying that they belong to the exact part of the cohomology of $\mathcal{B}_{\EuScript{Q}}$  \cite{Piguet:1995er}, \textit{i.e.} they appear only in $\Delta^{(-1)}$. Keeping this fact in mind, the most general structure allowed for $\Delta$ can be written as

\begin{eqnarray}
\Delta &=& \int \dd^4x\left[\frac{c_0}{4g^2}F^{a}_{\mu\nu}F^{a}_{\mu\nu}+c_1 (\partial_\mu (A^{h})^a_{\mu})(\partial_\nu (A^{h})^a_{\nu})+c_2 (\partial_\mu (A^{h})^a_{\nu})(\partial_\mu (A^{h})^a_{\nu})+c_3 f^{abc}(A^{h})^a_{\mu} (A^{h})^b_{\nu} \partial_\mu (A^{h})^c_{\nu}\right.\nonumber\\
&+&\left.\lambda^{abcd}(A^{h})^a_{\mu}(A^{h})^b_{\mu}(A^{h})^c_{\nu}(A^{h})^d_{\nu}+\hat{\mathcal{J}}^{a}_{\mu} \mathcal{O}^{a}_{\mu}(A,\xi)+J\mathcal{O}(A,\xi)+ c_4 (\partial_\mu \bar{\eta}^a+\Xi^{a}_{\mu})(\partial_\mu \eta^a)\right.\nonumber\\
&+&f^{abc}(\partial_\mu \bar{\eta}^a+\Xi^{a}_{\mu})\EuScript{P}^b_\mu(A,\xi)\eta^c+\left.c_5\frac{\theta}{2}J^2\right]\,,
\label{w21}
\end{eqnarray}
where  $(c_0,c_1,\ldots,c_5,\lambda)$ are arbitrary dimensionless  coefficients, while $\mathcal{O}^{a}_{\mu}(A,\xi)$, $\mathcal{O}(A,\xi)$ and $\EuScript{P}^b_\mu(A,\xi)$ are local expressions in $A^a_\mu$ and $\xi^a$ with ghost number zero and dimension one and two, respectively. \\\\In equation \eqref{w21} we have already taken into account the fact that, due to the Ward identity \eqref{wi4}, the variables $(\tau^a,\mathcal{J}^{a}_\mu)$ can enter the counterterm only through  the combination 
\begin{equation}
\hat{\mathcal{J}}^{a}_{\mu}=\mathcal{J}^{a}_{\mu}-\partial_\mu \tau^a \;.  \label{cb}
\end{equation}
Furthermore, from \eqref{w20}, one gets

\begin{eqnarray}
\mathcal{B}_{\EuScript{Q}}\mathcal{O}^{a}_{\mu}(A,\xi) &=& \EuScript{Q}\mathcal{O}^{a}_{\mu}(A,\xi) = s\mathcal{O}^{a}_{\mu}(A,\xi) = 0\,, \nonumber\\
\mathcal{B}_{\EuScript{Q}}\mathcal{O}(A,\xi) &=& \EuScript{Q}\mathcal{O}(A,\xi) = s\mathcal{O}(A,\xi) = 0\,,\nonumber\\
\mathcal{B}_{\EuScript{Q}}\EuScript{P}^a_\mu(A,\xi) &=& \EuScript{Q}\EuScript{P}^a_\mu(A,\xi) = s\EuScript{P}^a_\mu(A,\xi) = 0\,,
\label{w22}
\end{eqnarray}
implying the BRST-invariance of $\mathcal{O}^{a}_{\mu}(A,\xi)$, $\mathcal{O}(A,\xi)$ and $\EuScript{P}^a_\mu(A,\xi)$. In \cite{Fiorentini:2016rwx}, the general solution of eqs.\eqref{w22} was worked out\footnote{Although the explicit solution for $\EuScript{P}^a_\mu (A,\xi)$ was not shown in \cite{Fiorentini:2016rwx}, the reasoning is exactly the same as for $\mathcal{O}^{a}_{\mu}(A,\xi)$ and $\mathcal{O}(A,\xi)$.}, yielding

\begin{equation}
\mathcal{O}^{a}_{\mu} (A,\xi) = b_1 (A^{h})^a_{\mu}\,,
\label{w23}
\end{equation}
and

\begin{equation}
\mathcal{O} (A,\xi) = \frac{b_2}{2} (A^{h})^a_{\mu}(A^{h})^a_{\mu}\,,
\label{w24}
\end{equation}

\begin{equation}
\EuScript{P}^a_\mu(A,\xi) = b_3 (A^{h})^a_{\mu}\,,
\label{w24a}
\end{equation}
with $(b_1,b_2,b_3)$ free dimensionless parameters. As a consequence, the most general expression for $\Delta$ after the imposition of \eqref{w22} is expressed as

\begin{eqnarray}
\Delta &=& \int \dd^4x\left[\frac{c_0}{4g^2}F^{a}_{\mu\nu}F^{a}_{\mu\nu}+c_1 (\partial_\mu (A^{h})^a_{\mu})(\partial_\nu (A^{h})^a_{\nu})+c_2 (\partial_\mu (A^{h})^a_{\nu})(\partial_\mu (A^{h})^a_{\nu})+c_3 f^{abc}(A^{h})^a_{\mu} (A^{h})^b_{\nu} \partial_\mu (A^{h})^c_{\nu}\right.\nonumber\\
&+&\left.\lambda^{abcd}(A^{h})^a_{\mu}(A^{h})^b_{\mu}(A^{h})^c_{\nu}(A^{h})^d_{\nu}+b_1\hat{\mathcal{J}}^{a}_{\mu} (A^{h})^a_{\mu}+b_2\frac{J}{2}(A^{h})^a_{\mu}(A^{h})^a_{\mu}+ c_4 (\partial_\mu \bar{\eta}^a+\Xi^{a}_{\mu})(\partial_\mu \eta^a)\right.\nonumber\\
&+&\left.b_3 f^{abc}(\partial_\mu \bar{\eta}^a+\Xi^{a}_{\mu})(A^{h})^b_{\nu}\eta^c + c_5\frac{\theta}{2}J^2\right]\,.
\label{w25}
\end{eqnarray}
We should remark that, since the parameters $(\alpha,\chi)$ were introduced as a $\EuScript{Q}$-doublet, they do not enter in the non-trivial part of the  cohomology of $\EuScript{Q}$. As a consequence, these parameters do not appear in  $\Delta$. \\\\Before characterizing the exact part of the cohomology, {\it i.e.} $\Delta^{(-1)}$, one notices that if we set 
\begin{equation}
J=\tilde{J}=M=N=V=U=H=\chi=K=\mathcal{J}=\Xi=X=Y=\bar{X}=\bar{Y}=0
\label{sources_to_zero}
\end{equation}
in \eqref{ra20}, the resulting action is\footnote{Modulo the standard external BRST sources $(\Omega,L)$ terms.}

\begin{eqnarray}
\Sigma_{\mathrm{LCG}} &=& \frac{1}{4g^2}\int \dd^4x~F^{a}_{\mu\nu}F^{a}_{\mu\nu}+\int \dd^4x\left(ib^a \partial_\mu A^a_\mu+\frac{\alpha}{2}b^a b^a + \bar{c}^a\partial_\mu D^{ab}_\mu c^b\right)-\int \dd^4x\left(\bar{\varphi}^{ac}_{\mu}\EuScript{M}^{ab}(A^h){\varphi}^{bc}_{\mu}\right.\nonumber\\
&-&\left.\bar{\omega}^{ac}_\mu\EuScript{M}^{ab}(A^h)\omega^{bc}_\mu\right)+\int \dd^4x~\tau^a \partial_\mu (A^h)^a_\mu - \int \dd^4x~\bar{\eta}^a \EuScript{M}^{ab}(A^h)\eta^b\,,
\label{w26}
\end{eqnarray} 
which is the Yang-Mills gauge-fixed action in the linear covariant gauges with the addition of the following terms:

\begin{equation}
-\int \dd^4x\left(\bar{\varphi}^{ac}_{\mu}\EuScript{M}^{ab}(A^h){\varphi}^{bc}_{\mu}-\bar{\omega}^{ac}_\mu\EuScript{M}^{ab}(A^h)\omega^{bc}_\mu\right)+\int \dd^4x~\tau^a \partial_\mu (A^h)^a_\mu - \int \dd^4x~\bar{\eta}^a \EuScript{M}^{ab}(A^h)\eta^b\,.
\label{w27}
\end{equation}
However, upon integration over $(\bar{\varphi},\varphi,\bar{\omega},\omega)$ and $(\tau,\bar{\eta},\eta)$, the terms \eqref{w27} give rise to a unity. Therefore, correlation functions of the original  fields of the Faddeev-Popov quantization, \textit{i.e.} $(A,\bar{c},c,b)$, are the same as those computed with the standard Yang-Mills action in the linear covariant gauges \eqref{rgzlcg1}. From this observation, it follows that, in the limit \eqref{sources_to_zero}, the counterterm \eqref{w25} should reduce to the standard one in the Faddeev-Popov action in linear covariant gauges, see also \cite{Fiorentini:2016rwx}.  This gives 

\begin{equation}
c_1 = c_2 = c_3=0\,,\qquad c_4 = b_3\,,\qquad \lambda^{abcd}=0\,,
\label{w28}
\end{equation}
yielding 

\begin{eqnarray}
\Delta = \int \dd^4x\left[\frac{c_0}{4g^2}F^{a}_{\mu\nu}F^{a}_{\mu\nu} + b_1\hat{\mathcal{J}}^{a}_{\mu} (A^{h})^a_{\mu}+b_2\frac{J}{2}(A^{h})^a_{\mu}(A^{h})^a_{\mu}+ c_4 (\partial_\mu \bar{\eta}^a+\Xi^{a}_{\mu})D^{ab}_\mu(A^{h}) \eta^b+ c_5\frac{\theta}{2}J^2\right]\,.
\label{w29}
\end{eqnarray}
Finally, the Ward identity \eqref{w16}  imposes the following constraint

\begin{equation}
c_4 = - b_1\,. 
\label{w30}
\end{equation}
Let us proceed thus with the characterization of the trivial part of the cohomology of $\mathcal{B}_{\EuScript{Q}}$, \textit{i.e.} $\Delta^{(-1)}$. Keeping in mind that $\Delta^{(-1)}$ must have dimension bounded by four, be a local expression in the fields and sources and ghost number minus one, it follows that the most general term allowed by the constraint \eqref{w17} is
\begin{eqnarray}
\Delta^{(-1)}&=&\int \dd^{4}x\left[f^{ab}_{1}(\xi,\alpha)(\Omega^{a}_{\mu}+\partial_{\mu}\bar{c}^{a})A^{b}_{\mu}+f^{ab}_{2}(\xi,\alpha)c^{a}L^{b}+K^{a}f^{ab}(\xi,\alpha)\xi^{b}+\right.\nonumber\\
&-&\left.b_1\left(V^{ai}_{\mu}N^{ai}_{\mu}+V^{ai}_{\mu}D^{ab}_{\mu}(A^{h})\bar{\omega}^{bi}
+N^{ai}_{\mu}D^{ab}_{\mu}(A^{h})\varphi^{bi}
+(\partial_{\mu}\bar{\omega}^{ai})D^{ab}_{\mu}(A^{h})\varphi^{bi}\right)
\right]\,,
\label{w31}
\end{eqnarray}
with $f^{ab}_1 (\xi,\alpha)$, $f^{ab}_2 (\xi,\alpha)$ and $f^{ab}(\xi,\alpha)$ arbitrary functions of $\xi^a$ and $\alpha$. Invoking again the limit \eqref{sources_to_zero}, one is able to conclude that

\begin{equation}
f^{ab}_1 (\xi,\alpha) = \delta^{ab}d_1\,, \qquad f^{ab}_2 (\xi,\alpha) = \delta^{ab} d_2\,,
\label{w32}
\end{equation}
where $(d_1,d_2)$ are free parameters which might be $\alpha$-dependent. Acting with $\mathcal{B}_{\EuScript{Q}}$ on $\Delta^{(-1)}$ one obtains

\begin{eqnarray}
\mathcal{B}_{\EuScript{Q}}\Delta^{(-1)}&=&\int \dd^{4}x\,\bigg\{d_{1}\left(\frac{\delta\Sigma}{\delta A^{a}_{\mu}}+i\partial_{\mu}b^{a}\right)A^{a}_{\mu}-d_{1}(\Omega^{a}_{\mu}+\partial_{\mu}\bar{c}^{a})\frac{\delta\Sigma}{\delta \Omega^{a}_{\mu}}+d_{2}\left(\frac{\delta\Sigma}{\delta L^{a}}L^{a}+\frac{\delta\Sigma}{\delta c^{a}}c^{a}\right)+\frac{\delta\Sigma}{\delta \xi^{a}}f^{ab}(\xi)\xi^{b}\nonumber\\
&-& K^{b}\frac{\delta\Sigma}{\delta K^{a}}\left(\frac{\partial f^{bc}}{\partial \xi^{a}}\xi^{c}+f^{ba}(\xi)\right)+b_{1}f^{abc}(A^{h})^c_{\mu}\left(U^{ai}_{\mu}\bar{\omega}^{bi}+V^{ai}_{\mu}\bar{\varphi}^{bi}+M^{ai}_{\mu}{\varphi}^{bi} -N^{ai}_{\mu}{\omega}^{bi}-\omega^{ai}\partial_{\mu}\bar{\omega}^{bi}\right.\nonumber\\
&-&\left.\varphi^{ai}\partial_{\mu}\bar{\varphi}^{bi}\right)-b_1 \big[\,U^{ai}_{\mu}N^{ai}_{\mu}
+V^{ai}_{\mu}M^{ai}_{\mu}+U^{ai}_{\mu}\partial_{\mu}\bar{\omega}^{ai}+V^{ai}_{\mu}\partial_{\mu}\bar{\varphi}^{ai}+M^{ai}_{\mu}\partial_{\mu}{\varphi}^{ai}-N^{ai}_{\mu}\partial_{\mu}{\omega}^{ai}\nonumber\\
&+&(\partial_{\mu}\bar{\varphi}^{ai})\partial_{\mu}\varphi^{ai}-(\partial_{\mu}\bar{\omega}^{ai})\partial_{\mu}\omega^{ai}\,\big]
+\chi\frac{\partial d_2}{\partial\alpha}L^ac^a+\chi K^a\frac{\partial f^{ab}}{\partial\alpha}\xi^b+\chi\frac{\partial d_{1}}{\partial\alpha}(\Omega^{a}_{\mu}+\partial_{\mu}\bar{c}^{a})A^{a}_{\mu}\bigg\}\,.
\label{w33}
\end{eqnarray}
%$\clubsuit$ Moreover, from eq.\eqref{w33} we also deduce that $d_3$ $\clubsuit$ has to  be independent from $\alpha$. This property follows by realizing that, $\clubsuit$ in the limit   $(M,V,N,U)\to 0$, which amounts in fact to remove the Gribov horizon, it is apparent that the Zwanziger auxiliary fields $(\bar{\varphi}^{ai},\varphi^{ai}, \bar{\omega}^{ai},\omega^{ai})$ can be completely integrated out from  the action \eqref{ra20}, giving rise to a unity, a feature which is preserved by the counterterm when  the coefficient $d_3$ is precisely independent from $\alpha$. 
Therefore, for $\Sigma_{\mathrm{CT}}$, we get 
\begin{eqnarray}
\Sigma_{\mathrm{CT}} &=& \Delta + \mathcal{B}_{\EuScript{Q}}\Delta^{(-1)}\nonumber\\
&=& \int \dd^4x\left[\frac{c_0}{4g^2}F^{a}_{\mu\nu}F^{a}_{\mu\nu} + b_1\hat{\mathcal{J}}^{a}_{\mu} (A^{h})^a_{\mu}+b_2\frac{J}{2}(A^{h})^a_{\mu}(A^{h})^a_{\mu}-b_1 (\partial_\mu \bar{\eta}^a+\Xi^{a}_{\mu})D^{ab}_\mu(A^{h})\eta^b+ \frac{c_5}{2}J^2\right]\nonumber\\
&+&\int \dd^{4}x\,\bigg\{d_{1}\left(\frac{\delta\Sigma}{\delta A^{a}_{\mu}}+i\partial_{\mu}b^{a}\right)A^{a}_{\mu}-d_{1}(\Omega^{a}_{\mu}+\partial_{\mu}\bar{c}^{a})\frac{\delta\Sigma}{\delta \Omega^{a}_{\mu}}+d_{2}\left(\frac{\delta\Sigma}{\delta L^{a}}L^{a}+\frac{\delta\Sigma}{\delta c^{a}}c^{a}\right)+\frac{\delta\Sigma}{\delta \xi^{a}}f^{ab}(\xi)\xi^{b}\nonumber\\
&-& K^{b}\frac{\delta\Sigma}{\delta K^{a}}\left(\frac{\partial f^{bc}}{\partial \xi^{a}}\xi^{c}+f^{ba}(\xi)\right)+b_{1}f^{abc}(A^{h})^c_{\mu}\left(U^{ai}_{\mu}\bar{\omega}^{bi}+V^{ai}_{\mu}\bar{\varphi}^{bi}+M^{ai}_{\mu}{\varphi}^{bi} -N^{ai}_{\mu}{\omega}^{bi}-\omega^{ai}\partial_{\mu}\bar{\omega}^{bi}\right.\nonumber\\
&-&\left.\varphi^{ai}\partial_{\mu}\bar{\varphi}^{bi}\right)-b_1 \big[\,U^{ai}_{\mu}N^{ai}_{\mu}
+V^{ai}_{\mu}M^{ai}_{\mu}+U^{ai}_{\mu}\partial_{\mu}\bar{\omega}^{ai}+V^{ai}_{\mu}\partial_{\mu}\bar{\varphi}^{ai}+M^{ai}_{\mu}\partial_{\mu}{\varphi}^{ai}-N^{ai}_{\mu}\partial_{\mu}{\omega}^{ai}\nonumber\\
&+&(\partial_{\mu}\bar{\varphi}^{ai})\partial_{\mu}\varphi^{ai}-(\partial_{\mu}\bar{\omega}^{ai})\partial_{\mu}\omega^{ai}\,\big]
+\chi\frac{\partial d_2}{\partial\alpha}L^ac^a+\chi K^a\frac{\partial f^{ab}}{\partial\alpha}\xi^b+\chi\frac{\partial d_{1}}{\partial\alpha}(\Omega^{a}_{\mu}+\partial_{\mu}\bar{c}^{a})A^{a}_{\mu}\bigg\}\,.
\label{w34}
\end{eqnarray} 
Having determined the most general invariant local counterterm compatible with the symmetries of the theory given by \eqref{w34}, one should check the stability of the theory, namely, that \eqref{w34} can be reabsorbed in the original action \eqref{ra20} through a redefinition of the fields, parameters and sources. Before showing this, it is important to express \eqref{w34} in the parametric form, see \cite{Fiorentini:2016rwx}, a task which will be done in the following subsection.

\subsection{Parametric form of the counterterm}
%%%%%%%%%%%%%%%%%%%%%%%%%%%%%%%%%%%%%%%%%%%%%%%%%%%%

Having characterized the most general local invariant counterterm compatible with the Ward identities \eqref{w17}, let us proceed to prove that  it  can be re-absorbed in the original action $\Sigma$ by means of a suitable redefinitions of fields, sources and parameters. In this case, it turns out to be useful  to cast the counterterm in the parametric form, see \cite{Fiorentini:2016rwx}. To this end, we express the counterterm \eqref{w34} as\footnote{In the following, we have implemented the following redefinitions $(c_0,c_5,d_1,d_2)\to (a_0,b_3,a_1,a_2)$. Also, since we have already exploited  the dependence from  $\alpha$ of the counterterm, we can set $\chi=0$ from now on.}
\begin{equation}
\Sigma_{\mathrm{CT}} = \sum^{8}_{n=1} \Sigma^{\mathrm{CT}}_n
\label{w36}
\end{equation}
with
\begin{eqnarray}
\Sigma^{\mathrm{CT}}_1 &=& \frac{a_0}{4g^2}\int \dd^4x~F^{a}_{\mu\nu}F^{a}_{\mu\nu}\,,\nonumber\\
\Sigma^{\mathrm{CT}}_2 &=& b_1\int \dd^4x~\mathcal{J}^{a}_{\mu}A^{h,a}_\mu\,,\nonumber\\
\Sigma^{\mathrm{CT}}_3 &=& b_1\int \dd^4x~\left(\tau^a\partial_\mu A^{h,a}_\mu-(\partial_{\mu}\bar{\eta}^a+\Xi^{a}_{\mu})D^{ab}_{\mu}(A^h)\eta^b \right)\,,\nonumber\\
\Sigma^{\mathrm{CT}}_4 &=& \int \dd^4x~\left(b_2\frac{J}{2}A^{h,a}_\mu A^{h,a}_\mu+b_3\frac{\theta}{2}J^2\right)\,,\nonumber\\
\Sigma^{\mathrm{CT}}_5 &=& a_1\int \dd^4x\left(-ib^a \partial_\mu A^a_\mu\right)\,,\nonumber\\
\Sigma^{\mathrm{CT}}_6 &=& a_1 \int \dd^4x~\bar{c}^a\partial_\mu \frac{\delta\Sigma}{\delta\Omega^a_\mu}\,,\nonumber\\
\Sigma^{\mathrm{CT}}_7 &=& \int \dd^4x\left(a_1 A^a_\mu \frac{\delta \Sigma}{\delta A^a_\mu}-a_1\Omega^a_\mu\frac{\delta\Sigma}{\delta \Omega^a_\mu}+a_2L^a\frac{\delta\Sigma}{\delta L^a}+a_2 c^a\frac{\delta\Sigma}{\delta c^a}+f^{ab}(\xi)\xi^b\frac{\delta\Sigma}{\delta\xi^a}-K^a\frac{\delta\Sigma}{\delta K^a}\frac{\partial f^{bc}}{\partial \xi^a}\xi^c-K^b\frac{\delta\Sigma}{\delta K^a}f^{ba}(\xi)\right)\,,\nonumber\\
\Sigma^{\mathrm{CT}}_8 &=& b_1 \int \dd^4x\left(\bar{\varphi}^{ai}\EuScript{M}^{ab}(A^h)\varphi^{bi}-\bar{\omega}^{ai}\EuScript{M}^{ab}(A^h)\omega^{bi}+U^{ai}_\mu D^{ab}_\mu (A^h)\bar{\omega}^{bi}+V^{ai}D^{ab}_\mu (A^h)\bar{\varphi}^{bi}+M^{ai}_\mu D^{ab}_\mu (A^h)\varphi^{bi}\right.\nonumber\\
&-&\left.N^{ai}_\mu D^{ab}_\mu(A^h)\omega^{bi}+U^{ai}_\mu N^{ai}_\mu+V^{ai}_\mu M^{ai}_\mu\right)\,.
\label{w37}
\end{eqnarray}
One can employ eq.\eqref{ra22} to write

\begin{equation}
\Sigma^{\mathrm{CT}}_1 = -a_0 g^2\frac{\partial\Sigma}{\partial g^2}\,.
\label{w38}
\end{equation}
Also, one should notice that

\begin{eqnarray}
\frac{\delta\Sigma}{\delta \mathcal{J}^{a}_\mu}&=&A^{h,a}_\mu\,,\nonumber\\
\frac{\delta\Sigma}{\delta \tau^a} &=& \partial_\mu A^{h,a}_\mu\,,\nonumber\\
\frac{\delta\Sigma}{\delta\bar{\eta}^a}&=&\partial_{\mu}D^{ab}_{\mu}(A^{h})\eta^b\,,\nonumber\\
\frac{\delta\Sigma}{\delta{\eta}^a}&=&-D^{ab}_{\mu}(A^{h})\partial_{\mu}\bar{\eta}^b
-\bar{Y}^{abi}\varphi^{bi}+\bar{X}^{abi}\omega^{bi}+X^{i}\bar{\omega}^{ai}
-Y^{i}\bar{\varphi}^{bi}-D^{ab}_{\mu}(A^{h})\Xi^{b}_{\mu}\,,\nonumber\\
\frac{\delta\Sigma}{\delta X^{i}}&=&\eta^{a}\bar{\omega}^{ai}\,,\qquad
\frac{\delta\Sigma}{\delta Y^{i}}=\eta^{a}\bar{\varphi}^{ai}\,,\qquad
\frac{\delta\Sigma}{\delta \bar{X}^{abi}}=\eta^{a}{\omega}^{bi}\,,\qquad
\frac{\delta\Sigma}{\delta \bar{Y}^{abi}}=\eta^{a}{\varphi}^{bi}\,,\nonumber\\
\frac{\delta\Sigma}{\delta J} &=& \frac{1}{2}A^{h,a}_\mu A^{h,a}_\mu + \theta J\,,\nonumber\\
\theta \frac{\partial \Sigma}{\partial \theta} &=& \int \dd^4x~\frac{\theta}{2}J^2\,,
\label{w39}
\end{eqnarray}
which imply 

\begin{eqnarray}
\Sigma^{\mathrm{CT}}_2 &=& b_1\int \dd^4x~\mathcal{J}^{a}_{\mu}\frac{\delta\Sigma}{\delta \mathcal{J}^{a}_\mu}\,,\nonumber\\
\Sigma^{\mathrm{CT}}_3 &=& b_1\int \dd^4x\left[\tau^a \frac{\delta\Sigma}{\delta \tau^a} +\frac{1}{2}\left(\bar{\eta}^a \frac{\delta\Sigma}{\delta\bar{\eta}^a} + \eta^a \frac{\delta\Sigma}{\delta{\eta}^a}
+\Xi^a_{\mu} \frac{\delta\Sigma}{\delta{\Xi}^a_{\mu}} 
-X^{i}\frac{\delta\Sigma}{\delta X^{i}}
-Y^{i}\frac{\delta\Sigma}{\delta Y^{i}}
-\bar{X}^{abi}\frac{\delta\Sigma}{\delta\bar{X}^{abi}}
-\bar{Y}^{abi}\frac{\delta\Sigma}{\delta\bar{Y}^{abi}}\right) \right]\,,\nonumber\\
\Sigma^{\mathrm{CT}}_4 &=& b_2 \int \dd^4x~J\frac{\delta\Sigma}{\delta J}+(b_3-2b_2)\theta\frac{\partial\Sigma}{\partial \theta}\,.
\label{w40}
\end{eqnarray}
Concerning the term $\Sigma^{\mathrm{CT}}_5$, one recognizes that it  can be expressed in parametric form by taking into account that

\begin{equation}
\frac{\delta\Sigma}{\delta b^a}=i\partial_\mu A^a_\mu + \alpha b^a\,,\qquad \frac{\partial\Sigma}{\partial\alpha}=\frac{1}{2}b^a b^a\,.
\label{w41}
\end{equation}
Hence,

\begin{equation}
\Sigma^{\mathrm{CT}}_5 = -a_1 \int \dd^4x~b^a\frac{\delta\Sigma}{\delta b^a}+2a_1\alpha \frac{\partial\Sigma}{\partial \alpha}\,.
\label{w42}
\end{equation}
The term $\Sigma^{\mathrm{CT}}_6$ can be expressed in parametric form by using the fact that

\begin{equation}
\frac{\delta\Sigma}{\delta \bar{c}^a}=-\partial_\mu \frac{\delta \Sigma}{\delta \Omega^a_\mu}\,,
\label{w43}
\end{equation}
which yields

\begin{equation}
\Sigma^{\mathrm{CT}}_6 = -a_1\int \dd^4x~\bar{c}^a\frac{\delta \Sigma}{\delta \bar{c}^a}\,.
\label{w44}
\end{equation}
In order to write $\Sigma^{\mathrm{CT}}_8$ in parametric form, one should employ the relations

\begin{eqnarray}
\int \dd^4x~\bar{\varphi}^{ai}\frac{\delta\Sigma}{\delta\bar{\varphi}^{ai}}&=&\int \dd^4x\left[\bar{\varphi}^{ai}\partial^2\varphi^{ai}-f^{abc}A^{h,c}_\mu\varphi^{ai}\partial_\mu \bar{\varphi}^{bi}-V^{ai}_\mu \partial_\mu \bar{\varphi}^{ai}+f^{abc}A^{h,c}_\mu V^{ai}_\mu \bar{\varphi}^{bi}-\tilde{J}\bar{\varphi}^{ai}\varphi^{ai}+Y^{i}\eta^{a}\bar{\varphi}^{ai}\right]\,,\nonumber\\
\int \dd^4x~\bar{\omega}^{ai}\frac{\delta\Sigma}{\delta \bar{\omega}^{ai}}&=&\int \dd^4x\left[-\bar{\omega}^{ai}\partial^2\omega^{ai}+f^{abc}\bar{\omega}^{ai}\partial_\mu (A^{h,c}_\mu \omega^{bi})-\bar{\omega}^{ai}\partial_\mu U^{ai}_\mu - f^{bac}\bar{\omega}^{ai}U^{bi}_\mu A^{h,c}_\mu+\tilde{J}\bar{\omega}^{ai}\omega^{ai}\right.\nonumber\\
&+&\left.H\bar{\omega}^{ai}\varphi^{ai}+X^{i}\eta^{a}\bar{\omega}^{ai}\right]\,,\nonumber\\
\int \dd^4x~\varphi^{ai}\frac{\delta\Sigma}{\delta \varphi^{ai}} &=& \int \dd^4x\left[\varphi^{ai}\partial^2 \bar{\varphi}^{ai}+f^{abc}\varphi^{bi}(\partial_\mu \bar{\varphi}^{ai})A^{h,c}_\mu+\varphi^{ai}\partial_\mu M^{ai}_\mu+f^{abc}\varphi^{bi}M^{ai}_\mu A^{h,c}_\mu-\tilde{J}\bar{\varphi}^{ai}\varphi^{ai}\right.\nonumber\\
&+&\left.H\bar{\omega}^{ai}\varphi^{ai}+\bar{Y}^{abi}\eta^{a}\varphi^{bi}\right]\,,\nonumber\\
\int \dd^4x~\omega^{ai}\frac{\delta\Sigma}{\delta \omega^{ai}}&=& \int \dd^4x\left[\omega^{ai}\partial^2 \bar{\omega}^{ai}+f^{abc}\omega^{bi}(\partial_\mu \bar{\omega}^{ai})A^{h,c}_\mu+\omega^{ai}\partial_\mu N^{ai}_\mu + f^{abc}\omega^{bi}N^{ai}_\mu A^{h,c}_\mu - \tilde{J}\omega^{ai}\bar{\omega}^{ai}+\bar{X}^{abi}\eta^{a}\omega^{bi}\right]\,,\nonumber\\
\int \dd^4x~M^{ai}_\mu\frac{\delta\Sigma}{\delta M^{ai}_\mu} &=&\int \dd^4x\left[-M^{ai}_\mu D^{ab}_\mu (A^h) \varphi^{bi}-M^{ai}_\mu V^{ai}_\mu\right]\,,\nonumber\\
\int \dd^4x~V^{ai}_\mu\frac{\delta\Sigma}{\delta V^{ai}_\mu}&=&\int \dd^4x\left[-V^{ai}_\mu D^{ab}_\mu (A^h)\bar{\varphi}^{bi}-V^{ai}_\mu M^{ai}_\mu\right]\,,\nonumber\\
\int \dd^4x~N^{ai}_\mu \frac{\delta\Sigma}{\delta N^{ai}_\mu}&=&\int \dd^4x\left[N^{ai}_\mu D^{ab}_\mu (A^h)\omega^{bi}+N^{ai}_\mu U^{ai}_\mu\right]\,,\nonumber\\
\int \dd^4x~U^{ai}_\mu \frac{\delta\Sigma}{\delta U^{ai}_\mu}&=&\int \dd^4x\left[-U^{ai}_\mu D^{ab}_\mu (A^h)\bar{\omega}^{bi}-U^{ai}_\mu N^{ai}_\mu\right]\,,
\label{w45}
\end{eqnarray}
and

\begin{equation}
\frac{\delta\Sigma}{\delta \tilde{J}}=\bar{\omega}^{ai}\omega^{ai}-\bar{\varphi}^{ai}\varphi^{ai}\,,\qquad \frac{\delta\Sigma}{\delta H} = \bar{\omega}^{ai}\varphi^{ai}\,.
\label{w46}
\end{equation}
This entails that 

\begin{eqnarray}
\Sigma^{\mathrm{CT}}_8 &=& -\frac{b_1}{2} \int \dd^4x\left[\bar{\varphi}^{ai}\frac{\delta\Sigma}{\delta \bar{\varphi}^{ai}}+\varphi^{ai}\frac{\delta \Sigma}{\delta \varphi^{ai}}+\bar{\omega}^{ai}\frac{\delta\Sigma}{\delta \bar{\omega}^{ai}}+\omega^{ai}\frac{\delta\Sigma}{\delta\omega^{ai}}+M^{ai}_\mu \frac{\delta\Sigma}{\delta M^{ai}_\mu}+V^{ai}_\mu \frac{\delta\Sigma}{\delta V^{ai}_\mu}+N^{ai}_\mu \frac{\delta\Sigma}{\delta N^{ai}_\mu}+U^{ai}\frac{\delta\Sigma}{\delta U^{ai}_\mu}\right.\nonumber\\
&+&\left.2\tilde{J}\frac{\delta\Sigma}{\delta\tilde{J}}+2H\frac{\delta\Sigma}{\delta H}
-X^{i}\frac{\delta\Sigma}{\delta X^{i}}
-Y^{i}\frac{\delta\Sigma}{\delta Y^{i}}
-\bar{X}^{abi}\frac{\delta\Sigma}{\delta\bar{X}^{abi}}
-\bar{Y}^{abi}\frac{\delta\Sigma}{\delta\bar{Y}^{abi}}\right]\,.
\label{w47}
\end{eqnarray}
Consequently, the expression of the  counterterm $\Sigma^{\mathrm{CT}}$ in parametric form is

\begin{eqnarray}
\Sigma^{\mathrm{CT}} &=& -a_0 g^2\frac{\partial\Sigma}{\partial g^2}+b_1\int \dd^4x~\mathcal{J}^{a}_{\mu}\frac{\delta\Sigma}{\delta \mathcal{J}^{a}_\mu}+b_1\int \dd^4x\left[\tau^a \frac{\delta\Sigma}{\delta \tau^a} +\frac{1}{2}\left(\bar{\eta}^a \frac{\delta\Sigma}{\delta\bar{\eta}^a} + \eta^a \frac{\delta\Sigma}{\delta{\eta}^a} +\Xi^{a}_{\mu}\frac{\delta\Sigma}{\delta\Xi^{a}_{\mu}}\right) \right]+b_2 \int \dd^4x~J\frac{\delta\Sigma}{\delta J}\nonumber\\
&+&(b_3-2b_2)\theta\frac{\partial\Sigma}{\partial \theta}-a_1 \int \dd^4x~b^a\frac{\delta\Sigma}{\delta b^a}+2a_1\alpha \frac{\partial\Sigma}{\partial \alpha}-a_1\int \dd^4x~\bar{c}^a\frac{\delta \Sigma}{\delta \bar{c}^a}+\int \dd^4x\left[a_1 A^a_\mu \frac{\delta \Sigma}{\delta A^a_\mu}-a_1\Omega^a_\mu\frac{\delta\Sigma}{\delta \Omega^a_\mu}+a_2L^a\frac{\delta\Sigma}{\delta L^a}\right.\nonumber\\
&+&\left.a_2 c^a\frac{\delta\Sigma}{\delta c^a}+f^{ab}(\xi)\xi^b\frac{\delta\Sigma}{\delta\xi^a}-K^a\frac{\delta\Sigma}{\delta K^a}\left(\frac{\partial f^{bc}}{\partial \xi^a}\xi^c+f^{ba}(\xi)\right)\right]-\frac{b_1}{2} \int \dd^4x\left[\bar{\varphi}^{ai}\frac{\delta\Sigma}{\delta \bar{\varphi}^{ai}}+\varphi^{ai}\frac{\delta \Sigma}{\delta \varphi^{ai}}+\bar{\omega}^{ai}\frac{\delta\Sigma}{\delta \bar{\omega}^{ai}}\right.\nonumber\\
&+&\left.\omega^{ai}\frac{\delta\Sigma}{\delta\omega^{ai}}+M^{ai}_\mu \frac{\delta\Sigma}{\delta M^{ai}_\mu}+V^{ai}_\mu \frac{\delta\Sigma}{\delta V^{ai}_\mu}+N^{ai}_\mu \frac{\delta\Sigma}{\delta N^{ai}_\mu}+U^{ai}\frac{\delta\Sigma}{\delta U^{ai}_\mu}+2\tilde{J}\frac{\delta\Sigma}{\delta\tilde{J}}+2H\frac{\delta\Sigma}{\delta H}\right]\,,
\label{w48}
\end{eqnarray}
which can be immediately rewritten as 

\begin{equation}
\Sigma^{\mathrm{CT}} = \mathcal{R} \Sigma\,,
\label{w49}
\end{equation}
where $\mathcal{R}$  stands for the operator 

\begin{eqnarray}
\mathcal{R} &=& -a_0 g^2\frac{\partial}{\partial g^2}+b_1\int \dd^4x~\mathcal{J}^{a}_{\mu}\frac{\delta}{\delta \mathcal{J}^{a}_\mu}+b_1\int \dd^4x\left[\tau^a \frac{\delta}{\delta \tau^a} +\frac{1}{2}\left(\bar{\eta}^a \frac{\delta}{\delta\bar{\eta}^a} + \eta^a \frac{\delta}{\delta{\eta}^a} +\Xi^{a}_{\mu}\frac{\delta}{\delta\Xi^{a}_{\mu}}\right) \right]+b_2 \int \dd^4x~J\frac{\delta}{\delta J}\nonumber\\
&+&(b_3-2b_2)\theta\frac{\partial}{\partial \theta}-a_1 \int \dd^4x~b^a\frac{\delta}{\delta b^a}+2a_1\alpha \frac{\partial}{\partial \alpha}-a_1\int \dd^4x~\bar{c}^a\frac{\delta}{\delta \bar{c}^a}+\int \dd^4x\left[a_1 A^a_\mu \frac{\delta}{\delta A^a_\mu}-a_1\Omega^a_\mu\frac{\delta}{\delta \Omega^a_\mu}+a_2L^a\frac{\delta}{\delta L^a}\right.\nonumber\\
&+&\left.a_2 c^a\frac{\delta}{\delta c^a}+f^{ab}(\xi)\xi^b\frac{\delta}{\delta\xi^a}-K^a\left(\frac{\partial f^{bc}}{\partial \xi^a}\xi^c+f^{ba}(\xi)\right)\frac{\delta}{\delta K^a}\right]-\frac{b_1}{2} \int \dd^4x\left[\bar{\varphi}^{ai}\frac{\delta}{\delta \bar{\varphi}^{ai}}+\varphi^{ai}\frac{\delta}{\delta \varphi^{ai}}+\bar{\omega}^{ai}\frac{\delta}{\delta \bar{\omega}^{ai}}\right.\nonumber\\
&+&\left.\omega^{ai}\frac{\delta}{\delta\omega^{ai}}+M^{ai}_\mu \frac{\delta}{\delta M^{ai}_\mu}+V^{ai}_\mu \frac{\delta}{\delta V^{ai}_\mu}+N^{ai}_\mu \frac{\delta}{\delta N^{ai}_\mu}+U^{ai}\frac{\delta}{\delta U^{ai}_\mu}+2\tilde{J}\frac{\delta}{\delta\tilde{J}}+2H\frac{\delta}{\delta H}\right]\,.
\label{w50}
\end{eqnarray}
Expression \eqref{w49} turns out to be quite useful for the analysis of the satbility of the starting action $\Sigma$, as  addressed in the next subsection. 

%%%%%%%%%%%%%%%%%%%%%%%%%%%%%%%%%%%%%%%%%%%%%%%%%%%%
\subsection{Stability of the action \texorpdfstring{$\Sigma$}{TEXT}}
%%%%%%%%%%%%%%%%%%%%%%%%%%%%%%%%%%%%%%%%%%%%%%%%%%%%

In order to end  the algebraic proof of the renormalizability of \eqref{ra20}, one should prove that the counterterm \eqref{w49} can be re-absorbed into the initial action $\Sigma$ through  a suitable redefinition of fields, sources and parameters. The determination of those redefinitions is made very easy once one knows the counterterm written in its parametric form, as in \eqref{w48}. If the counterterm  \eqref{w48} can be re-abosrbed in the starting action, then, to the first order in the parameter expansion $\epsilon$, the following relation should hold  \cite{Piguet:1995er}, {\it i.e.} 

\begin{equation}
\Sigma[\Phi_0] = \Sigma[\Phi] + \epsilon \Sigma^{\mathrm{CT}}[\Phi] + O(\epsilon) \,,   
\label{w51}
\end{equation}
where $\Phi$ stands for all fields, sources and parameters of the theory. From \eqref{w49}, one concludes that

\begin{equation}
\Sigma[\Phi_0] = \Sigma[\Phi] + \epsilon \mathcal{R}\Sigma [\Phi] + O(\epsilon) \,,   
\label{w52}
\end{equation}
and due to the form of $\mathcal{R}$, it is easy to see that

\begin{equation}
\Phi_0 = (1+\epsilon\mathcal{R})\Phi\,.
\label{w53}
\end{equation}
The fields, sources and parameters are redefined thus according to 

\begin{eqnarray}
A_0 &=& Z^{1/2}_A A\,,\qquad b_0 = Z^{1/2}_b b\,,\qquad c_0 = Z^{1/2}_c c\,, \qquad \bar{c}_0 = Z^{1/2}_{\bar{c}} \bar{c}\,, \nonumber\\
\xi^{a}_0&=& Z^{ab}_{\xi}(\xi)\xi^b\,,\qquad \tau_0 = Z^{1/2}_\tau \tau\,,\qquad \eta_0 = Z^{1/2}_\eta \eta\,,\qquad \bar{\eta}_0 = Z^{1/2}_{\bar{\eta}}\bar{\eta}\,,\nonumber\\
\bar{\varphi}_0&=&Z^{1/2}_{\bar{\varphi}}\bar{\varphi}\,,\qquad \varphi_0 = Z^{1/2}_\varphi \varphi\,, \qquad \bar{\omega}_0 = Z^{1/2}_{\bar{\omega}}\bar{\omega}\,,\qquad \omega_0=Z^{1/2}_{\omega}\omega\,,\nonumber\\
\Omega_0&=&Z_{\Omega}\Omega\,,\qquad L_0=Z_L L\,,\qquad K^{a}_0 = Z^{ab}_{K}(\xi)K^b\,,\qquad \mathcal{J}_0 = Z_{\mathcal{J}}\mathcal{J}\,,\nonumber\\
J_0&=&Z_J J\,,\qquad \tilde{J}_0 = Z_{\tilde{J}}\tilde{J}\,,\qquad H_0 = Z_H H\,,\qquad g_0=Z_g g\,,\nonumber\\
\alpha_0&=&Z_{\alpha}\alpha\,,\qquad \theta_0 = Z_\theta \theta\,,\qquad M_0 = Z_M M\,,\qquad V_0 = Z_V V\,,\nonumber\\
N_0&=& Z_N N\,,\qquad U_0 =  Z_U U\,,\qquad \Xi_0=Z_{\Xi}\,\Xi\,,\qquad X_0=Z_{X}X\,,\nonumber\\
Y_0&=&Z_{Y} Y\,,\qquad \bar{X}_{0}=Z_{\bar{X}}\bar{X}\,,\qquad \bar{Y}_{0}=Z_{\bar{Y}}\bar{Y}\,.
\label{w54}
\end{eqnarray}
with 

\begin{eqnarray}
Z^{1/2}_A &=& 1+\epsilon a_1\,,\qquad Z^{1/2}_c = 1+\epsilon a_2\,,\qquad Z_g = 1-\epsilon\frac{a_0}{2}\,,\qquad Z^{1/2}_\tau = 1+\epsilon b_1\,,\qquad Z_J = 1+\epsilon b_2\nonumber\\
Z_\theta &=& 1+\epsilon (b_3-2b_2)\,,\qquad Z^{ab}_{\xi}(\xi) = \delta^{ab}+\epsilon f^{ab}(\xi)\,,\qquad Z^{ab}_K (\xi)= \delta^{ab}-\epsilon\left(f^{ba}(\xi)+\frac{\partial f^{bc}}{\partial \xi^a}\xi^c\right)\,.
\label{w55}
\end{eqnarray}
For the other fields, sources and parameters, the following relations hold

\begin{eqnarray}
Z^{1/2}_A &=& Z^{-1}_{\Omega} = Z^{-1/2}_{\bar{c}} = Z^{-1/2}_b = Z^{1/2}_\alpha\,,\nonumber\\
Z^{1/2}_\tau &=& Z_{\bar{\eta}} = Z_{\eta} =Z^{2}_{\Xi}= Z_{\mathcal{J}}\,,\nonumber\\
Z^{1/2}_c &=& Z_L\,,\qquad Z_X=Z_Y=Z_{\bar{X}}=Z_{\bar{Y}}=1\,,
\label{w56}
\end{eqnarray}
while for those fields and sources introduced to implement the Gribov horizon one has

\begin{eqnarray}
Z^{-1/4}_\tau = Z^{1/2}_{\bar{\varphi}} = Z^{1/2}_{\varphi} = Z^{1/2}_{\bar{\omega}} = Z^{1/2}_{\omega} = Z_M = Z_V = Z_N = Z_U = Z^{1/2}_{\tilde{J}}=Z^{1/2}_H\,.   
\label{w57}
\end{eqnarray}
We see thus that, under an appropriate redefinition of fields, sources and parameters as described in eq.(\ref{w55}), \eqref{w56} and \eqref{w57}, the most general local invariant counterterm compatible with the Ward identities can be re-absorbed in the classical action \eqref{ra20}. Hence, by the algebraic renormalization framework \cite{Piguet:1995er}, the theory is renormalizable at all orders in perturbation theory. It is important to emphasize that, due to the fact that the Stueckelberg field is dimensionless, the associated  renormalization factors, $(Z^{ab}_{\xi}(\xi), Z^{ab}_K (\xi))$ are nonlinear in $\xi^a$, a feature typical of dimensionless fields, see also \cite{Fiorentini:2016rwx}. \\\\An interesting reamark is that the renormalization factor of the Gribov parameter $\gamma^2$ is not an independent quantity of the theory, being expressed in terms of other renormalization factors. In fact,  taking the physical limit of the sources, see \eqref{ra9} and \eqref{ra21}, it turns out that

\begin{equation}
Z_{\gamma^2} = Z^{-1/2}_{\mathcal{J}} = Z^{1/2}_{\mathcal{J}}Z^{-1/2}_{\tau} =  1-\frac{\epsilon}{2} b_1\,.
\label{w58}
\end{equation}
In addition, we also have 
\begin{equation}
Z_{\mathcal{J}}Z^{1/2}_{\varphi}Z^{1/2}_{\bar{\varphi}}=1\,,\qquad \qquad Z_{\mathcal{J}}Z^{1/2}_{\omega}Z^{1/2}_{\bar{\omega}}=1\,,
\label{w59}
\end{equation}
which express the nonrenormalization properties of the vertices $(A^h \phi {\bar \phi})$ and $(A^h \omega {\bar \omega})$, already noticed in  \cite{Capri:2016aif} where the study of the  renormalizability of the Refined Gribov-Zwanziger action in the linear covariant gauges in the approximation $A^h \approx A^{T}$, with $A^T$ the transverse component of the gauge field, was studied. \\\\In conclusion, the action \eqref{ra20} takes into account the existence of  infinitesimal Gribov copies in the linear covariant gauges in a local, BRST invariant and renormalizable way. 

%%%%%%%%%%%%%%%%%%%%%%%%%%%%%%%%%%%%%%%%%%%%%%%%%%%%%%%%%%%%%%
\section{Conclusions}
%%%%%%%%%%%%%%%%%%%%%%%%%%%%%%%%%%%%%%%%%%%%%%%%%%%%%%%%%%%%%%

In the Landau gauge, dealing with Gribov copies has brought non-trivial infrared effects which might be related to the confinement of gluons, see \cite{Vandersickel:2011zc,Greensite:2011zz,Brambilla:2014jmp,Deur:2016tte}, a fact that  has raised  the investigation of such effects in other gauges. In particular,  the issue of the  Gribov copies in the linear covariant gauges has been object of intense research in the last years, see \cite{Capri:2015ixa,Capri:2016aqq,Capri:2015nzw,Capri:2016gut,Sobreiro:2005vn,Capri:2015pja,Capri:2016aif,Capri:2017abz}. In particular, a local and BRST invariant action which takes into account the existence of Gribov copies in the linear covariant gauges was proposed in \cite{Capri:2016aqq}, within the Gribov-Zwanziger framework. In the present work, we have been pursuing the study of the action constructed in \cite{Capri:2016aqq}, by proving its renormalizability to all orders in perturbation theory. This provides  a consistent framework in order to perform loop computations, a subject which is under current investigation. The present work can be naturally  extended to the study of the renormalizability of the Refined-Gribov-Zwanziger action in the presence  of non-perturbative matter coupling, as devised in \cite{Capri:2014bsa}. The cases of the the maximal Abelian and Curci-Ferrari gauges, see \cite{Capri:2015pfa,Pereira:2016fpn}, can be addressed as well.

%%%%%%%%%%%%%%%%%%%%%%%%%%%%%%%%%%%%%%%%%%%%%%%%%%%%%%%%%%%%%%
\section*{Acknowledgements}
%%%%%%%%%%%%%%%%%%%%%%%%%%%%%%%%%%%%%%%%%%%%%%%%%%%%%%%%%%%%%%

The authors are thankful to U. Reinosa, J. Serreau, M. Tissier and N. Wschebor for discussions. The Conselho Nacional de Desenvolvimento Cient\'{i}fico e Tecnol\'{o}gico (CNPq-Brazil) and The Coordena\c c\~ao de Aperfei\c coamento de Pessoal de N\'ivel Superior (CAPES) are acknowledged for support.

\appendix

\section{Remarks on the localization of the BRST-invariant RGZ action} \label{localRGZap} 
%%%%%%%%%%%%%%%%%%%%%%%%%%%%%%%%%%%%%%%%%%%%%%%%%%%%%%%%%%%%%%%%%%%%%%%%

In this appendix, we explicitly show that the BRST invariant local formulation of the Refined Gribov-Zwanziger action \eqref{intro19} in terms of $A^h_\mu$ in the Landau gauge is equivalent to the original construction presented in \cite{Dudal:2008sp}. We begin with the BRST invariant Refined Gribov-Zwanziger action in the Landau gauge expressed as

\begin{eqnarray}
S^{L}_{\mathrm{RGZ}} &=& S_{\mathrm{YM}} + \int \dd^4x\left(ib^a \partial_\mu A^a_\mu + \bar{c}^a\partial_\mu D^{ab}_{\mu}c^b\right)+\int\dd^4x\left[\varphi^{ac}_{\mu}\partial_\nu D^{ab}_\nu (A^h)\varphi^{bc}_\mu - \bar{\omega}^{ac}_\mu \partial_\nu D^{ab}_{\nu}(A^h)\omega^{bc}_\mu - \gamma^2 f^{abc}A^{h,a}_\mu (\varphi+\bar{\varphi})^{bc}_\mu\right]\nonumber\\
&+& \frac{m^2}{2}\int \dd^4x~A^{h,a}_\mu A^{h,a}_\mu - M^2\int \dd^4x\left(\bar{\varphi}^{ab}_\mu\varphi^{ab}_\mu-\bar{\omega}^{ab}_\mu\omega^{ab}_\mu\right)  \int \dd^4x~\tau^{a}\partial_\mu A^{h,a}_\mu + \int \dd^4x~\bar{\eta}^a \partial_\mu D^{ab}_{\mu}(A^h)\eta^b\,. 
\label{ap1}
\end{eqnarray}
The partition function is written as 

\begin{equation}
\EuScript{Z} = \int \left[\EuScript{D}\Phi\right]\mathrm{e}^{-S^{L}_{\mathrm{RGZ}}}\,,
\label{ap2}
\end{equation}
where $\Phi = \left\{A,b,\bar{c},c,\bar{\varphi},\varphi,\bar{\omega},\omega,\xi,\tau,\bar{\eta},\eta\right\}$. Integrating out the fields $(b,\tau,\bar{\eta},\eta)$ one obtains

\begin{equation}
\EuScript{Z} = \int \left[\EuScript{D}\tilde{\Phi}\right]\delta\left(\partial_\mu A^a_\mu\right)\delta\left(\partial_\mu A^{h,a}_\mu\right)\mathrm{det}\left(-\partial_\mu D^{ab}_{\mu}(A^h)\right)\mathrm{e}^{-\int \dd^4x\left(\ldots\right)}\,,
\label{ap3}
\end{equation} 
with $\left(\ldots\right)$ a shorthand notation for the remaining terms in the action \eqref{ap1} and $\tilde{\Phi}=\left\{A,\bar{c},c,\bar{\varphi},\varphi,\bar{\omega},\omega,\xi\right\}$. In order to deal with the delta function $ \delta\left(\partial_\mu A^{h,a}_\mu\right)$ imposing the transversality condition $\partial_\mu A^{h,a}_\mu=0$, we make use of  

\begin{equation}
\delta(f(x)) = \frac{\delta (x-x_0)}{|f'(x_0)|}\,,
\label{ap4}
\end{equation}
with $f(x_0)=0$. Of course, this relation holds if $f'(x_0)$ exists and is non-vanishing. It is possible to construct an iterative solution for $\partial_\mu A^{h,a}_\mu=0$, as described in Appendix~A of \cite{Capri:2015ixa}. Such a solution $\xi_0$ is expressed as

\begin{equation}
\xi_0 = \frac{1}{\partial^2}\partial_\mu A_\mu + \frac{ig}{\partial^2}\left[\partial_\mu A_\mu , \frac{\partial_\nu A_\nu}{\partial^2}\right]+\frac{ig}{\partial^2}\left[A_\mu,\partial_\mu \frac{\partial_\nu A_\nu}{\partial^2}\right]+\frac{ig}{2}\frac{1}{\partial^2}\left[\frac{\partial_\mu A_\mu}{\partial^2},\partial_\nu A_\nu\right]+\mathcal{O}(A^3)\,,
\label{ap5}
\end{equation}
where we have employed the matrix notation of Appendix~A of \cite{Capri:2015ixa}. The important feature of \eqref{ap5} is that all terms  always contain   the divergence of the gauge field, {\it i.e.} $\partial_\mu A^a_\mu$. \\\\Hence, the analogue of \eqref{ap4} is

\begin{equation}
\delta(\partial_\mu A^{h,a}_\mu) = \frac{\delta (\xi - \xi_0)}{\mathrm{det}\left(-\partial_\mu D^{ab}_\mu (A^h)\right)}\,,
\label{ap6}
\end{equation}
where,  due to the restriction of the  domain of integration in the functional integral  to the Gribov region, we have taken into account that  ${\mathrm{det}\left(-\partial_\mu D^{ab}_\mu (A^h)\right)}>0$. Thus, plugging \eqref{ap6} into \eqref{ap3} yields

\begin{equation}
\EuScript{Z} = \int \left[\EuScript{D}\tilde{\Phi}\right]\delta\left(\partial_\mu A^a_\mu\right)\delta(\xi-\xi_0)\mathrm{e}^{-\int \dd^4x\left(\ldots\right)}\,.
\label{ap7}
\end{equation} 
Moreover,  one easily sees from \eqref{ap5} that,  due to the presence of $\delta (\partial_\mu A^a_\mu)$, it follows that $\xi_0 = 0$. Therefore, 

\begin{equation}
\EuScript{Z} = \int \left[\EuScript{D}\tilde{\Phi}\right]\delta\left(\partial_\mu A^a_\mu\right)\delta(\xi)\mathrm{e}^{-\int \dd^4x\left(\ldots\right)}\,.
\label{ap8}
\end{equation} 
Finally, reminding that $A^h_\mu  =( h^\dagger A_\mu h +\frac{i}{g}h^\dagger \partial_\mu h )$ with $ h = \mathrm{e}^{ig\xi^a T^a}$,  integration over $\xi$ gives $A^h_\mu  =( h^\dagger A_\mu h +\frac{i}{g}h^\dagger \partial_\mu h )  \to A$, so that  the original formulation of the Refined Gribov-Zwanziger action in the Landau gauge, as presented in \cite{Dudal:2008sp}, has been recovered.

\end{document}